# A Recent Survey of the Advancements in Deep Learning Techniques for Monkeypox Disease Detection

Saddam Hussain Khan[1]*, Rashid Iqbal[1], Saeeda Naz[1]

[1]Artificial Intelligence Lab, Department of Computer Systems Engineering, University of Engineering and Applied Sciences (UEAS), Swat, Pakistan

**Email:** saddamhkhan@ueas.edu.pk

## Abstract

Monkeypox (MPox) is a zoonotic infectious disease induced by the MPox Virus, part of the poxviridae orthopoxvirus group initially discovered in Africa and gained global attention in mid-2022 with cases reported outside endemic areas. Symptoms include headaches, chills, fever, smallpox, measles, and chickenpox-like skin manifestations and the WHO officially announced MPox as a global public health pandemic, in July 2022. Timely diagnosis is imperative for assessing disease severity, conducting clinical evaluations, and determining suitable treatment plans. Traditionally, PCR testing of skin lesions is considered a benchmark for the primary diagnosis by WHO, with symptom management as the primary treatment and antiviral drugs like tecovirimat for severe cases. However, manual analysis within hospitals poses a substantial challenge including the substantial burden on healthcare professionals, limited facilities, availability and fatigue among doctors, and human error during public health emergencies, particularly in the case of epidemics and pandemics. Therefore, this survey paper provides an extensive and efficient analysis of deep learning (DL) methods for the automatic detection of MPox in skin lesion images. These DL techniques are broadly grouped into categories, including deep CNN, Deep CNNs ensemble, deep hybrid learning, the newly developed, and Vision transformer for diagnosing MPox. Moreover, this study offers a systematic exploration of the evolutionary progression of DL techniques and identifies, and addresses limitations in previous methods while highlighting the valuable contributions and innovation. Additionally, the paper addresses benchmark datasets and their collection from various authentic sources, pre-processing techniques, and evaluation metrics. The survey also briefly delves into emerging concepts, identifies research gaps, limitations, and applications, and outlines challenges in the diagnosis process. This survey furnishes valuable insights into the prospective areas of DL innovative ideas and is anticipated to serve as a path for researchers.

## 1. Introduction

Infectious diseases, resulting from various pathogens such as microbes, parasites, germs, and fungi, pose significant threats to both humans and animals. These pathogens can be transmitted through various means, including biofluid exposure, contact with contaminated objects, close interaction with infected individuals, and ingesting food or water that has been tainted or polluted [1]. Many contagious diseases, such as common colds and influenza, can be transmitted through the air and the environment and are typically treated with antimicrobial drugs and other medications. The spectrum of infectious ailments is diverse while symptoms often accompany infectious diseases, some can be asymptomatic [2]. Moreover, zoonotic diseases, which transfer from infected animals to humans and then spread among humans, also fall under the category of infectious diseases [3]. Historical instances, like the 1918-19 Spanish flu, resulted in an estimated 50 million global fatalities [4]. Similarly, Monkeypox (MPox), a zoonotic infectious disease caused by the Orthopoxvirus, shares a genetic affinity with both horsepox and smallpox (SPox) [5]. In addition, Figure 1 depicts the symptoms, while Table 1 provides information on various aspects of infectious diseases.

### 1.1. Monkeypox

MPox viruses all belong to the Orthopox taxonomic group within the poxvirus family [5]. Its primary transmission occurs through contact with Baboons and Beavers, and then frequently transmitted between individuals [6]. The virus was first discovered in a Danish monkey during the year 1958 [7]. In 1970, an initial individual MPox case was documented in the Congo during campaigns to eliminate SPox [8]. MPox predominantly emerges in sub-Saharan Africa, affecting communities near tropical rainforests [9]. The virus spreads through physical contact, animal scratches or bites, airborne particles, or mucosal membranes in facial features [13]. Initial symptoms contain pyrexia, muscle pain, and tiredness, which later progress to the development of distinctive skin lesions [14]. MPox cases are on the rise, cases have been identified outside of Africa in 2022, with 94 nations and approximately 83,424 patients by December 21, 2022 [10]–[12]. This surge has led to increasing anxiety and discussion on online platforms [13]. At present, no targeted therapy or cure exists for MPox as per Centers for Disease Control and Prevention guidelines, but two oral drugs, Brin cidofovir and Tecovirimat, originally developed for SPox, are being repurposed for MPox treatment [14]. Vaccination remains the ultimate solution, although Food and Drug Administration-approved vaccines for MPox are not administered in the United States, using SPox vaccines instead [15].

### 1.2. Infectious Diseases Beyond MPox

Infectious diseases have continued to impact both developed and developing nations in the twenty-first century. Vaccine development poses considerable challenges and exerts adverse economic effects on many countries [16]. Recent updates from the World Health Organization (WHO) have brought to light several infectious disease outbreaks. Notably, in 2014, there was the emergence of the measles epidemic at a California amusement park, followed by a Zika epidemic that impacted millions of people across 60 regions of the world. Here is a list of infectious diseases: COVID-19, Idiopathic Hepatitis, Measles, Dengue, and Chickenpox (CPox) [17].

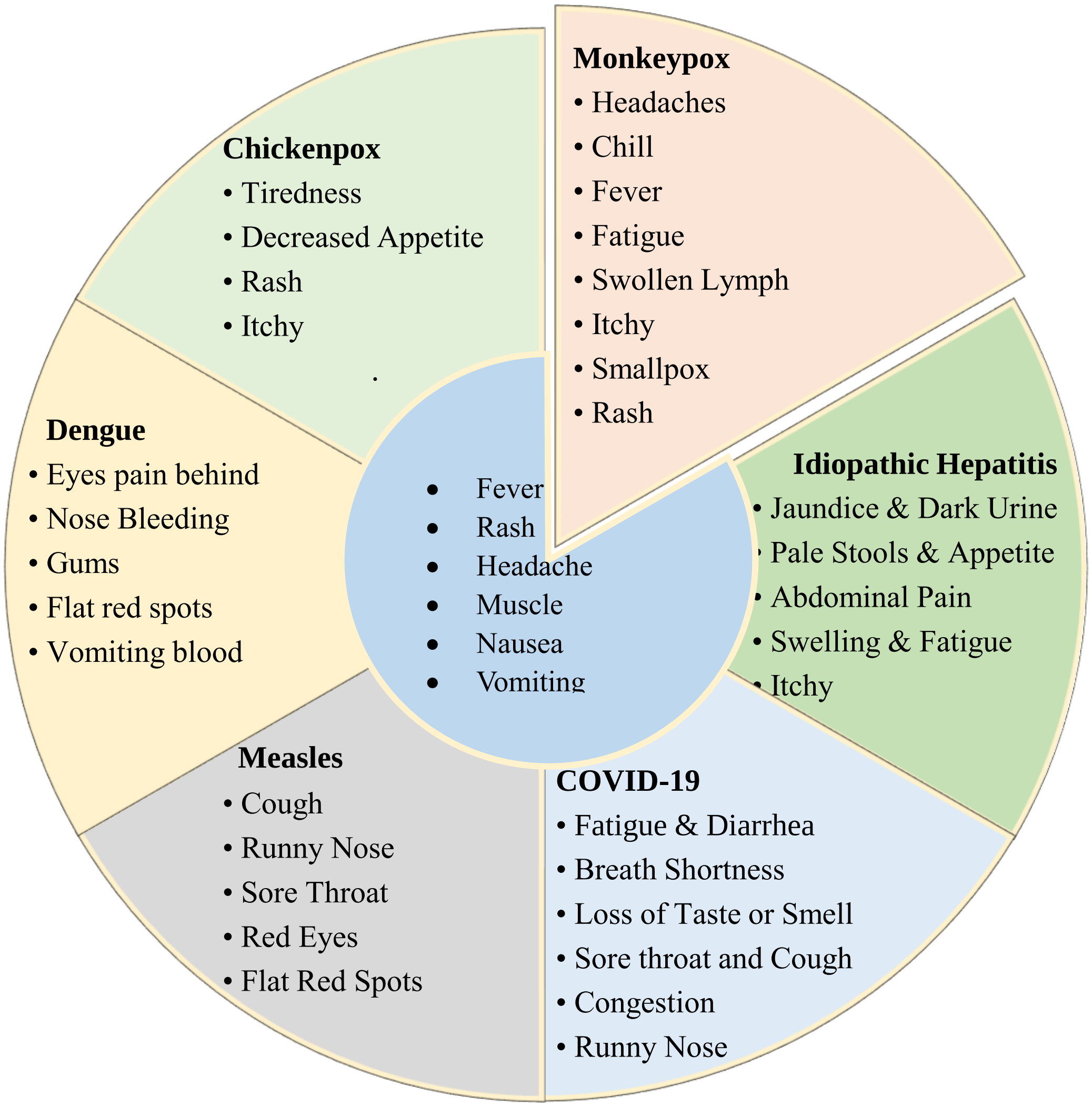

Figure 1: Comparison of Symptoms: MPox, CPox, Measles, Dengue, COVID-19, and Idiopathic Hepatitis

#### 1.2.1. COVID-19

Severe Acute Respiratory Syndrome Coronavirus-2 SARS-CoV-2, which is a part of the coronavirus group, is the cause of the COVID-19 outbreak that began in December 2019 with the first identified case in Wuhan City, China. WHO officially declared the global health crisis on 11th March 2020. The virus has been rapidly spread worldwide, with staggering statistics as of November 22, 2023, reporting more than 772 million verified instances and more than 7 million fatalities [18], [19]. The impact of the pandemic has been profound, affecting public health, economies, employment, and mental well-being. Vulnerable groups, including the elderly, youth, women, people of color, and the immunocompromised, have been disproportionately affected.

The virus has an incubation period of roughly two weeks, and infected individuals can remain asymptomatic. Diagnosis primarily depends on PCR tests using nasal swabs [20]. Treatment involves supplemental oxygen therapy, antibiotics to prevent microbial and yeast coinfections, and corticosteroids to reduce mortality risk for hospitalized patients. Public health measures, such as social restrictions, travel limitations, and social distancing, have significantly reduced transmission. Vaccination has played a pivotal role in decreasing mortality rates and the severity of symptomatic cases.

### 1.2.2. Idiopathic Hepatitis

Idiopathic Hepatitis, as defined by the WHO, is characterized by acute hepatitis in children aged 16 or younger, excluding hepatitis A, B, C, D, and E. It is distinguished by elevated serum transaminase levels exceeding 500 IU/L. Adenovirus, primarily transmitted through respiratory droplets, is frequently detected, with Adenovirus type 41 being the most prevalent strain [21]. Immunological liver injury induced by adenovirus is a suspected causative factor. Notably, approximately 76.60% affect youngsters who are in the age group of 5 years or younger. Up to July 8, 2022, approximately 1,010 cases and 22 deaths have been documented in 35 nations, in addition to 90 instances and 4 fatalities documented. Some common manifestations include fatigue, elevated body temperature, respiratory issues, diarrhea, nausea, and abdominal pain. Some patients experience severe outcomes, necessitating intensive care or liver transplantation. Diagnosis relies on laboratory testing and radiological examinations like ultrasounds of the abdomen or MRIs. Supportive care constitutes the mainstay of treatment, though complications may arise, requiring liver transplantation. Preventative measures comprise Proper hand cleanliness, maintaining physical distancing, Maintaining adequate airflow, masking, consumption of Purified water, surface disinfection, and safe food handling [21].

### 1.2.3. Measles

Measles is a very infectious illness that is attributed to a single RNA virus found within the Morbillivirus genus within the Paramyxoviridae family. It primarily spreads through direct contact and airborne droplets. The incubation period typically spans 10 to 14 days. Measles initially targets the respiratory system and subsequently disseminates systemically within the body, resulting in immune suppression. In 2019, this viral scourge affected roughly 870,000 individuals worldwide, resulting in 208,000 reported deaths, predominantly among children aged 5 or younger. While still endemic in certain underprivileged regions, vaccination campaigns have contributed to a notable 73% reduction in cases from 2000 to 2018. Swift recognition and vaccination efforts are pivotal in mitigating its impact[22].

Clinical symptoms of measles generally commence with elevated body temperature, followed by nasal discharge, cough, irritated eyes, and the appearance of characteristic ‘Koplik spots’ inside the cheeks. After 3 to 5 days, a maculopapular rash emerges, starting on the head and progressively spreading across the body. Complications associated with measles encompass diarrhea, ear infections, lung inflammation, brain inflammation, premature delivery, below-average birth weight, and potential fatality. The risk is significantly higher for babies, adults above 20, expectant mothers, individuals with weakened immune systems, and individuals lacking vitamin A. Notably, patients can transmit the virus starting four days prior to the rash's appearance and continuing for four days afterward. Diagnosis relies on clinical findings and laboratory tests. While no specific antiviral treatment is available, supportive care, nutritional support, hydration, vitamin A supplementation, and antibiotics for secondary infections are part of the treatment regimen. The prevention of measles hinges on vaccination, with specific age-based guidelines recommending two doses of the MMR vaccine [23].

#### 1.2.4. Dengue

Dengue, caused by the Dengue Virus with four subtypes, is transmitted by infected Aedes mosquitoes and through human-to-mosquito transmission. Notably, Dengue sub-neutralizes antibodies through antibody-dependent enhancement, subsequently facilitating infection in monocytes, macrophages, and dendritic cells. This infectious disease is prevalent in tropical and subtropical regions, affecting over half of the global population, as reported by the WHO. Typically, Dengue manifests as a mild disease, but it can escalate to severe forms, with mortality rates ranging from 2% to 5%. Prevention strategies encompass the management of mosquito breeding sites, personal protection, public awareness campaigns, and robust surveillance.

Clinical manifestations of Dengue cover a spectrum from asymptomatic cases to mild or severe diseases, experiencing flu-like signs emerging within four to ten days post-infection. Severe dengue, occurring in approximately 1 out of 20 patients, is characterized by clinical features like intense headache, elevated body temperature, queasiness, throwing up, muscle discomfort, spinal discomfort, and skin eruption. Additionally, severe cases may lead to complications, including plasma leakage, severe bleeding, and organ involvement. Diagnosis of Dengue makes use of methods such as RT-PCR and ELISA. While there is no specific antiviral treatment, a vaccine has demonstrated efficacy in individuals with prior dengue infection. Furthermore, ongoing clinical trials hold the promise of additional vaccine options in the future. Routine vaccination efforts have already resulted in a significant reduction in deaths related to measles.

#### 1.2.5. Chickenpox

Chickenpox (CPox), commonly known as varicella, is a highly contagious and sometimes severe infection resulting from the varicella-zoster virus (VZV), a part of the herpesvirus group. Notably, VZV is characterized by a single serotype, and humans are its exclusive known reservoir. CPox primarily affects children, individuals over 50, and those with weakened immune systems. Most children acquire VZV immunity by age 15, either naturally or through vaccination. Shingles, which usually affects adults over 60, occurs when the dormant virus reactivates in 10–30% of cases. Vaccination can prevent both CPox and shingles, but the schedules and doses differ for children and adults.

However, many countries do not implement CPox vaccination because of several reasons. First, severe complications are rare, affecting less than 1% of cases. Second, childhood vaccination might reduce natural exposure in adults, leading to less immune-boosting and more reactivation. Third, low vaccine coverage could shift the infection age to older groups with higher risks. Fourth, natural infection may provide longer immunity than vaccination.

Some countries opt for vaccination to prevent serious cases, such as hospitalizations, which amount to 4 million annually in the United States. The main challenge is the presumed impact of CPox vaccination on immune-boosting, which refers to the reinforcement of VZV-specific immunity, mainly cellular immunity. T-cell-mediated immunity protects against reactivation. Theoretical models predict an increase in shingles incidence after introducing CPox vaccination, but empirical data are lacking.

Traditionally, the WHO has regarded PCR testing of skin lesions as a standard for initial diagnosis, coupled with symptom management as the primary form of treatment. In severe cases, antiviral drugs such as tecovirimat are prescribed. However, PCR facilities are restricted and susceptible to a false-negative rate influenced by viral load and sampling strategy. To address this limitation, medical imaging is employed as an auxiliary detection tool to mitigate the false-negative rate associated with PCR. However, the visual analysis of medical images poses a substantial burden on radiologists, impacting their overall performance. Therefore, classical ML methods were utilized in medical image analysis due to their ability to operate effectively with limited training data and computational resources. Nevertheless, the current landscape demands advancements in ML and DL algorithms for the analysis of complex and abundant medical image data.

Table 1: Overview of Infectious Diseases.

| **Disease** | **Origin** | **Source** | **Diagnosis** | **Transmission** | **Affected organs** | **Survival** | **Damages** | **Countries** |
|---|---|---|---|---|---|---|---|---|
| MPox | Democratic Republic Congo, 1970 | Rodents, and monkeys | PCR, Sequencing, Serological tests, | Contact with bodily fluids, respiratory droplets, or contaminated | Skin, lymph nodes, mouth and throat, lungs, liver, kidneys, and brain | 2 -4 weeks | Recent 2 years 0.092 million cases | 116 |
| Measles | Boston, USA, 1757 | Humans | Specific IgM antibodies in the blood or saliva | Respiratory droplets or direct contact with nasal or throat secretions | Skin, Ear, Eyes, pneumonia, and bronchitis | 2 hours | 1980-2022 47 million cases | 187 |
| CPox | London, England, 1767 | Humans | PCR, ELISA, or DFA | Respiratory droplets or direct contact with fluid from the blisters | Skin, Ear, lung, brain, and blindness | 2-4 weeks | --- | Worldwide |
| Dengue | Philippines 1950s. | Mosquitoes, Aedes aegypti & Aedes albopictus | viral RNA, or specific antibodies in the blood or serum | Bite of infected mosquitoes | Damage to blood and lymph vessels | 2–7 days | 5.2 million in 2019 | 100 |
| COVID-19 | Wuhan, China, 2019 | Bats, pangolins, or other animals | viral RNA by RT-PCR or antigen by RDTs | Respiratory droplets, aerosols, or direct contact with mucous membranes | Lungs, respiratory, Cardiovascular system, Kidneys, Brain, Pancreas, Liver, Immune. | 4 weeks | 772 million | 229 |

### 1.3. AI's Role in Infectious MPox Disease Detection

Artificial Intelligence (AI) serves a versatile purpose in the field of infectious ailments and healthcare, spanning applications in diagnosis, epidemiology, treatment, and the exploration of antimicrobial drug resistance. It particularly excels in managing the vast datasets generated during disease outbreaks, thereby enhancing our ability to predict potential pandemic hotspots. A notable instance of AI's effectiveness is its pivotal role in tracking and analyzing the global spread of COVID-19 across diverse geographical regions [24].

#### 1.3.1. Machine Learning-based MPox Detection

Several studies have effectively harnessed AI, especially in the field of Machine Learning (ML) to tackle communicable illness. Molecular diagnostic procedures and ML for detecting sepsis, demonstrate its potential in analyzing extensive clinical datasets and offering support for treatment decisions [25]. Bayesian ML method has been employed to identify effective drug components against the Ebola virus, paving the way for effective drug discovery [26]. ML-based approaches have been delved into for Lyme disease diagnosis, showcasing the remarkable precision of their sparse support vector machine (SVM) classification methods [27]. The integration of High-Resolution Melt Analysis (HRM) is a quick method for post-PCR DNA analysis. SVMs, renowned for their classification capabilities, have proven highly accurate in the identification of isolated bacteria. This amalgamation leverages the sensitivity of HRM for detecting genetic variations and the classification prowess of SVM, ultimately elevating diagnostic precision [28]. Furthermore, the application of techniques such as SVM with leave-one-out cross-validation (LOOCV) has achieved a remarkable 100% precision rate in the identification of isolated bacteria [29]. These studies exemplify the diverse applications of ML and AI in the realm of communicable illness.

A wide array of ML methods, like SVM, KNN, naive Bayes, random forests (RF), K-means, and dendrogram analysis, are extensively utilized in the study of communicable diseases [30]. Moreover, ML's contributions extend to the realm of microorganism detection and subsequent infection surveillance, playing a vital role in advancing the understanding of infectious diseases and their management.

#### 1.3.2. Deep Learning for MPox Detection

Deep learning (DL) methods have demonstrated their effectiveness in automatically diagnosing various viral and bacterial diseases, especially MPox diagnosis using skin lesion images. DL has been instrumental in various applications for diagnosing infectious illnesses. Therefore, the use of artificial neural networks (ANN) DL architecture is described for quick and accurate assessment of microscopic images related to various types of measles [31]. Moreover, the DL-based approach is described to predict Zika virus infections, achieving a notable accuracy rate with a multilayer perceptron neural network (MLPNN) [32]. Additionally, the DL-based approach is focused on detecting the cell division stage in Uterine Leiomyosarcoma (ULMS), enhancing the precision and effectiveness of diagnosis [33].

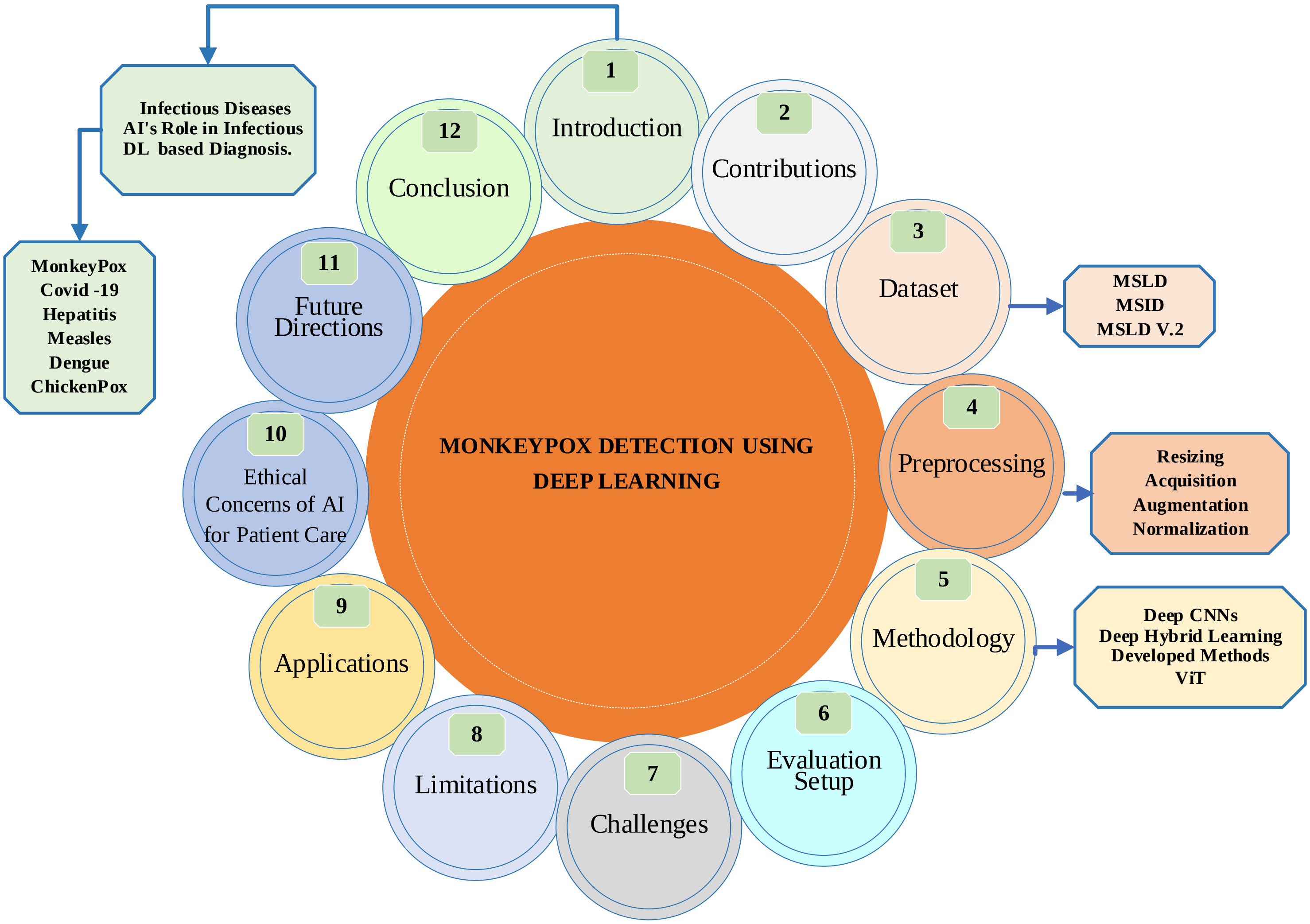

Figure 2: Layout of the Proposed Survey Article.

In the realm of medical imaging, DL-driven techniques have been exemplified by the deployment of Convolutional Neural Networks (CNNs) to analyze COVID-19 patient imaging datasets, effectively categorizing patients based on disease severity levels [34]. Notably, Rajawat et al. developed the COVID Net, a multilayer CNN model trained on a combined image dataset, showcasing its efficiency in classifying both type and disease severity [35]. The hierarchical structure of CNNs inherently yields exceptional precision in image data analysis.

DL's impact extends to large-scale data analysis in the healthcare sector, proving instrumental in the diagnosis of both communicable and non-communicable diseases [36]. It plays a crucial role in assisting disease outbreak analysis. DL interventions utilize various data sources, including radiological ultrasound for tuberculosis and microscopy for malaria [37]. Furthermore, DL effectively utilizes data from sources such as PCR and MRI in the diagnosis of diseases like SARS and COVID-19.

The paper is organized as follows. In Section 2, we explore the contribution of the survey, and Section 3 delves into the use of standard datasets for DL-based MPox virus detection. In Section 4, we explore preprocessing techniques, encompassing contrast enhancement, image resizing, and data augmentation. Moving on to Section 5, we elucidate the methodology, covering deep CNNs, deep CNN ensemble, hybrid learning, the developed DL methods, and Vision Transformer (ViT). Then, the spotlight shifts toward the evaluation and experimental setup in Section 6. In Section 7, we addressed various challenges within the field, associated with dataset collection, the implementation of DL techniques, and the medical impact and approvals required. Following this, Section 8 dissects the limitations associated with the methodologies discussed. Section 9 shifts the focus towards recent applications of DL in medical practice, particularly in the realms of detection and treatment. Moreover, we provide valuable information about DL and web-based resources pertinent to medical practice, catering to the growing importance of online information. In Section 10, the paper thoroughly explores the ethical considerations associated with the research, addressing concerns such as patient privacy and responsible data usage. Section 11 outlines the direction for future research, emphasizing the importance of quality-oriented investigations and DL advancements in healthcare. Finally, Section 12 serves as the concluding part of the paper, summarizing the key findings and contributions. To provide readers with a quick overview, Figure 2 outlines the paper's flow.

## 2. Contributions

This research aims to elucidate the methodologies employed by researchers in MPox disease detection, encompassing picture pre-processing, feature extraction, and detection techniques. Additionally, it sheds light on the effective utilization of DL algorithms in MPox disease detection, with a primary focus on research conducted up to October 2023. DL algorithms offer a more efficient approach to MPox disease detection, involving fewer steps compared to traditional ML methods. The generalized steps are embedded within their block diagram for MPox detection, as depicted in Figure 3 This review article stands as a valuable resource for researchers in this field, providing a comprehensive overview of isolated DL models for practical applications. The systematic contribution of our survey article is as follows:

1. Comprehensive analysis of DL techniques for MPox detection in skin lesion images. Moreover, the study presents a systematic survey encompassing infectious diseases, their symptoms, causes, associated damages, strategies for diagnosis, and treatment planning.
2. Categorization of DL methods, including deep CNN, Deep CNNs ensemble and hybrid learning, newly developed approaches, and ViT for MPox detection.

3. This study additionally offers a systematic and comprehensive exploration of DL techniques, addressing limitations in previous approaches and highlighting the invaluable contributions, impact, and innovations within DL that enhance detection capabilities.
4. In-depth exploration of benchmark datasets sourced from authentic repositories that contain extensive information on datasets used by various researchers. Moreover, pre-processing strategies and evaluation metrics have been discussed.
5. We have identified and discussed the emerging concepts, research gaps, diagnostic process challenges, limitations, and practical applications for MPox detection.
6. Offering valuable insights into promising avenues for future DL research, architectural innovation ideas, and development of potential applications as a guide for the research community.

This structured approach enhances the readability and applicability of research findings, enabling researchers to efficiently access and implement the knowledge presented in this survey. It simplifies the understanding and navigation of the research landscape in MPox disease detection, rendering it a valuable resource for those within the field.

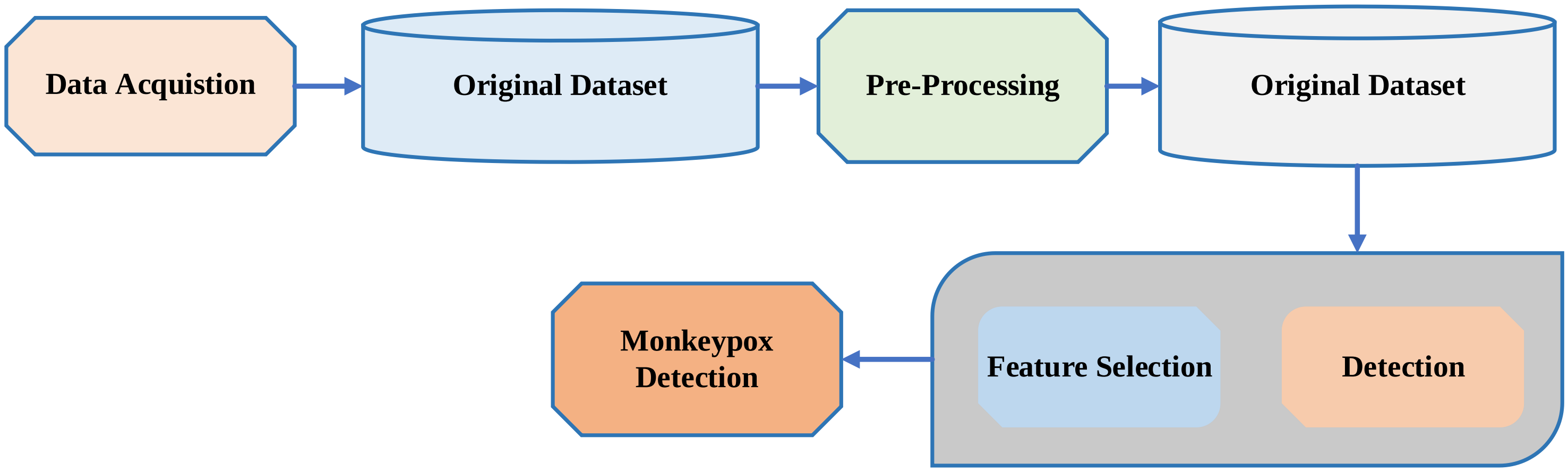

Figure 3: Steps of MPox disease detection using DL.

## 3. Standard Datasets

Nine distinct datasets were gathered from diverse sources, including Kaggle, GitHub, and hospitals, and employed in multiple studies aimed at the identification and detection of MPox. The dataset details are displayed in Figure 4 which includes images of MPox, measles, CPox, and normal cases, as shown in Figure 5. In 2023, the Computational Science & Engineering Department at the Islamic University Kushtia, Bangladesh, introduced a constructed skin image dataset named "**MPox Skin Image Dataset (MSID)**" designed for MPox detection. The dataset has four classes: MPox, Measles, CPox, and Healthy, and is published on Kaggle at [38] and on Mendeley at [39] to expedite the research and treatment activities. Moreover, "Dataset of MPox Skin Lesions**: Version2 (MSLD v2)"** (2023) encompasses six distinct categories, namely MPox, CPox, Measles, Cowpox, Hand-foot-mouth disease (HFMD), and Normal. The dataset has been acquired in collaboration with experienced dermatologists and represents an enhanced iteration of the MSLD. Furthermore, a web application prototype has been crafted employing this dataset. Additionally, the "**Dataset of MPox Skin Lesions (MSLD)**" was meticulously curated through the collection & processing of photos from multiple web sources using web data extraction. These sources included online platforms, news websites, and publicly available case records. The dataset is classified into two main groups: MPox and others. Its primary purpose is to facilitate the discrimination of MPox cases from similar non-MPox cases. An enhanced version of this dataset, known as "MSLD v2.0," was subsequently released in 2023. Furthermore, a new synthetic dataset, named the "**MPox PATIENTS Dataset**" was created based on a study published by the BMJ.

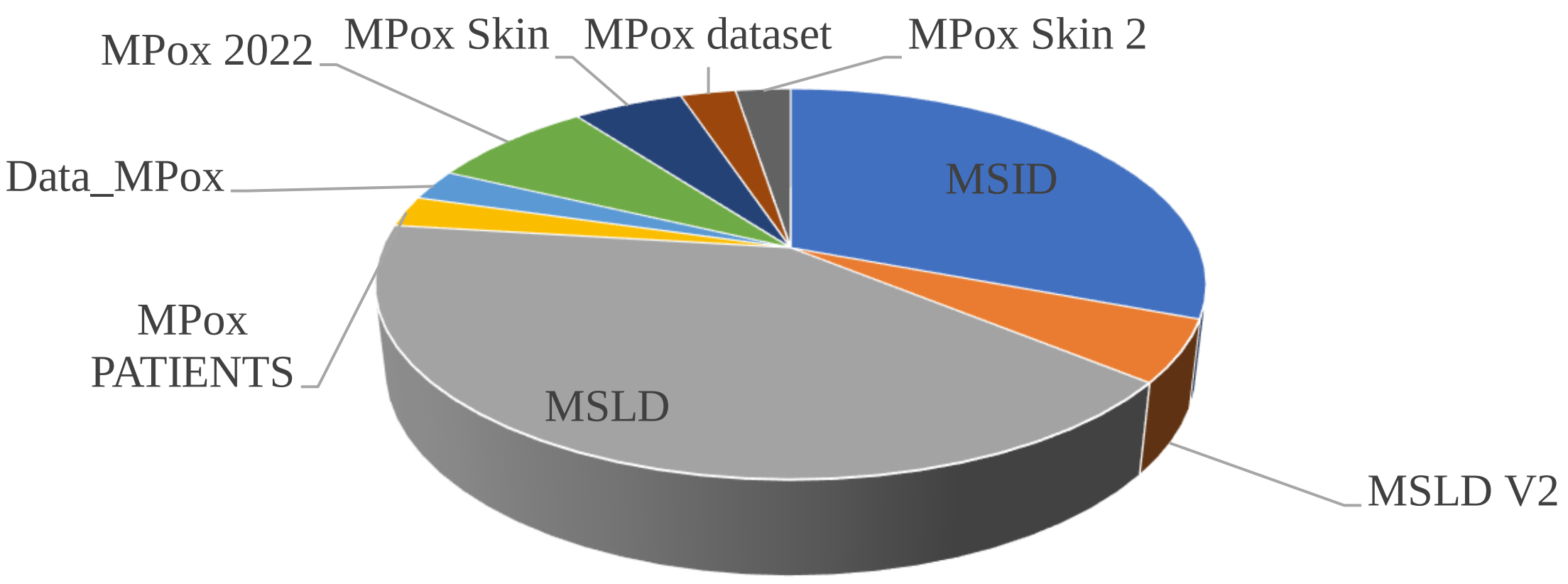

Figure 4: Monkeypox (MPox) Dataset Overview

Table 2: MPox Datasets Details

| Source | Dataset's Name | Dimension | Class | Link |
|---|---|---|---|---|
| **Kaggle** <br> **Mendeley Data** | MPox Skin Img Dataset (MSID) | MPox (279), CPox (107) Measles (91), Normal (293), Total (770) | 4 | www.kaggle.com/datasets/dipuiucse/monkeypoxskinimagedataset. <br> data.mendeley.com/datasets/r9bfpnvyxr/6. |
| **Kaggle** | MPox Skin Lesions: (MSLD v2) | MPox (284), CPox (75), Measles (55), Cowpox (66), HFMD (161), Healthy (114): Total (755) | 6 | www.kaggle.com/datasets/joydippaul/mpox-skin-lesion-dataset-version-20-msld-v20. |
| **Github** | MPox-dataset-2022 | MPox (43), MPox_Aug (587): CPox(11) : Normal (54), Normal_Aug (552): Measles (17), Measles_Aug (286): Total (1550) | 4 | github.com/mahsan2/monkeypox-dataset-2022. |
| **Kaggle** | MPox Skin Lesions | MPox (102), Others (126), Total (228) | 2 | www.kaggle.com/datasets/nafin59/monkeypox-skin-lesion-dataset. |
| **Github** | MPoxSkin dataset | MPox (100), Others (100), Healthy (100), Total (300) | 3 | github.com/mjdominguez/monkeypoxSkinimage. |
| **Social Media** | MPox | MPox (356), Non-MPox (345) | 2 | 2209.02415.pdf (arxiv.org) |
| **Kaggle** | MPox PATIENTS Dataset. | Total patients (25000) | 2 | www.kaggle.com/datasets/muhammad4hmed/monkeypox-patients-dataset. |
| **Roboflow** | MPox Skin Dataset 2 | Positive (MPox Lesions):1390, Negative (Other Diseases):708, Total (2098) | 2 | app.roboflow.com/ds/uHWnw424Sk?key=w8YJKfcD2i. |
| **Kaggle** | MPox-image-dataset | MPox (264), CPox (100), Measles (80), Normal (215), Total (659) | 4 | www.kaggle.com/datasets/sachinkumar413/MP-image-dataset. |
| **Kaggle** | Data_MPox | MPox (45), MPox Aug (533): Normal (25), Normal Aug (298): Roseola (25), Roseola Aug (297): Skarlet: (22), Skarlet_Aug (263): Total (1508) | 4 | www.kaggle.com/datasets/ahmadnasayrah/data-monkeypox. |

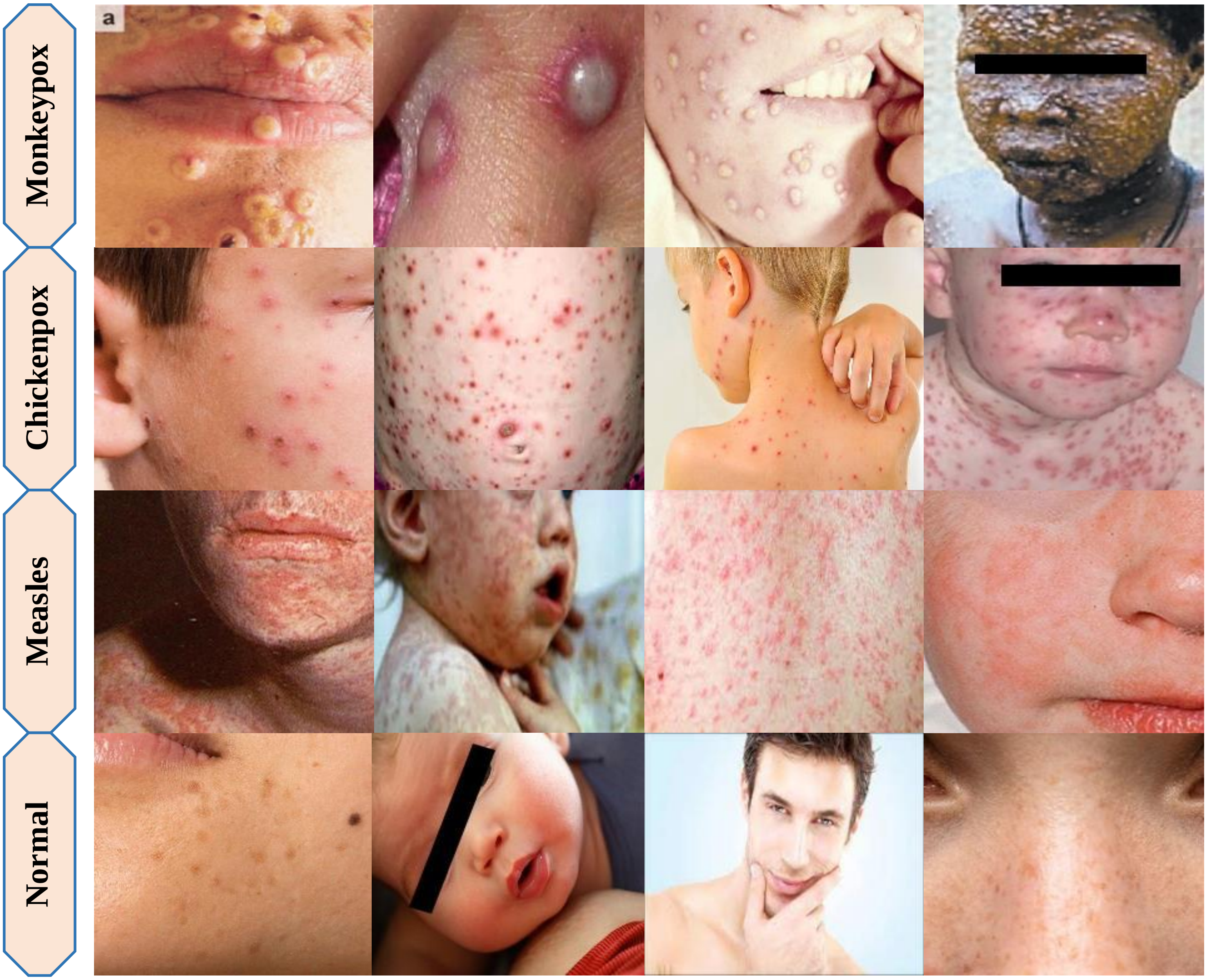

Figure 5: Examples of healthy, MPox, CPox, and Measles-affected samples.

This dataset comprises CSV records of patients along with their relevant characteristics and a designated variable signifying the existence or non-existence of MPox. Notably, this dataset includes Boolean and categorical features. Moreover, a dataset named "**Data_MPox**" was curated, encompassing four distinct classes: MPox, Roseola, Scarlet Fever, and Normal. In the same vein, **MPox-dataset-2022** was curated using sources from Google and the CDC, and it is accessible via www.cdc.gov.This dataset is thoughtfully structured into three specific categories: MPox, Measles, and Normal. It is made accessible on GitHub [40]. It's worth mentioning that this dataset has been specifically designed for educational and research purposes. Deploying it for real-world patient detection is strongly discouraged, as the dataset is not associated with the recent MPox outbreak and has not obtained approval from an ethical committee. Likewise, the "**MPoxSkin dataset**" was created at the University of Seville, Spain. This dataset comprises images sourced from various datasets and trusted webpages, such as the CDC. These images were then resized to square sections measuring 224x224 pixels. The dataset is categorized into three distinct groups: MPox, Healthy, and Other Skin Damages. Its primary objective is to enhance the early identification and assessment of MPox. In addition,

In 2022, Moise et al. [41] introduced an "**MPox dataset**". This dataset includes images sourced from publicly available medical literature, social media, and journalistic sources. It is categorized into a pair of classes: MPox and non-MPox. The selection criteria for the images in this dataset were based on confirmed diagnoses, with all MPox images belonging to clade II and included only if they were officially recorded as verified through testing. Besides this, the "**MPox Skin Dataset 2**" was established, featuring images collected from the web and made available on robo.ai [42]. This dataset is categorized into two main groups: positive (MPox) and negative (similar diseases). Additionally, for testing purposes, it includes MPox images from the Kaggle Lesion Dataset. All the dataset details are presented in Table 2.

## 4. Preprocessing

Skin lesion images play a pivotal role in medical research and the diagnosis of various infectious diseases using DL. However, the process of preparing the dataset faces artifacts and variations that must be addressed before the skin lesions can be effectively analyzed. Preprocessing skin images is a crucial step, significantly enhancing the performance of computational systems and aiding both human and machine analysis. Common preprocessing techniques, such as Contrast Enhancement, Normalization, Data Augmentation (DA), Image Resizing, and Dataset Acquisition, are frequently employed to enhance dataset quality, enhance robustness, and improve model generalization.

### 4.1. Contrast Enhancement and Normalization

The Synthetic Minority Over-Sampling Method (SMOM) has been utilized to address data imbalance and enhance the reliability of the dataset in combination with the Edited Nearest Neighbors (ENN) method. This process resulted in a well-balanced dataset with reduced sensitivity to noise. In various research studies, image contrast is enhanced by applying standardized dimensions to accommodate various pixel sizes, enhancing the model's ability to process data seamlessly. To mitigate issues arising from variations in lighting and contrast, a meticulous normalization filter was thoughtfully applied to the images. Furthermore, the pixel values underwent rescaling, transitioning from their original scale of 0 to 255 to a more practical range of 0 to 1.

### 4.2. Image Resizing

One valuable technique in the preprocessing of images is image scaling, which aims to optimize the compatibility of input images with the DL model. It is crucial to emphasize that the resizing step occurs prior to feeding the images into the models, effectively simulating real-time processing.

### 4.3. Data Augmentation

Data Augmentation (DA) is a key method used to increase the dataset's size by including slightly altered versions of the available data or by generating new data from the existing dataset. This process encompasses various transformation techniques, including flipping, rotation (0–360 degrees), padding, re-scaling, translation (shifting the image along the X and Y directions), cropping, zooming, adjusting brightness and contrast, applying grayscale, introducing noise, and random erasing. Employing these methods enables the creation of a more extensive and diverse dataset for training DL models, thereby facilitating the generation of new images with subtle variations from the originals.

DA techniques are normally employed to improve the generalization of models. Generalization pertains to the model's capacity to improve in tasks involving data it has not encountered previously. It's essential to reduce the distinction between the anticipated risk of a function and the empirical risk, as this ensures better generalization. The generalization gap, represented as the difference between expected and empirical risks, can be reduced through various techniques. DA has emerged as a prominent approach for augmenting data quantity, particularly in domains where obtaining precise data may be a formidable task, as in the context of MPox research. Beyond its role in dataset expansion, DA serves as a regularizer, playing a pivotal role in mitigating overfitting issues during the training of ML models.

### 4.4. Dataset Acquisition

Initially, there was a lack of publicly available datasets specifically dedicated to MPox images. To bridge this gap, the researchers compiled datasets from open-source web resources and patient records, categorizing the images into positive (MPox) and negative (non-MPox) classes. To expand the dataset, DA techniques were applied.

## 5. Methodology

The methodology section of the survey encompasses DL techniques for diagnosing MPoxV. The diagnostic approach encompasses four distinct DL experimental configurations: (1) the utilization of pre-existing deep CNNs (explained in **Section 5.1**), (2) the implementation of deep CNN ensemble within a hybrid learning framework (**Section 5.2**), (3) the integration of newly developed techniques (**Section 5.3**), and (4) leveraging the ViT (**Section 5.4**). The detailed workflow of the MPox detection framework is outlined below and is also visually represented in Figure 6.

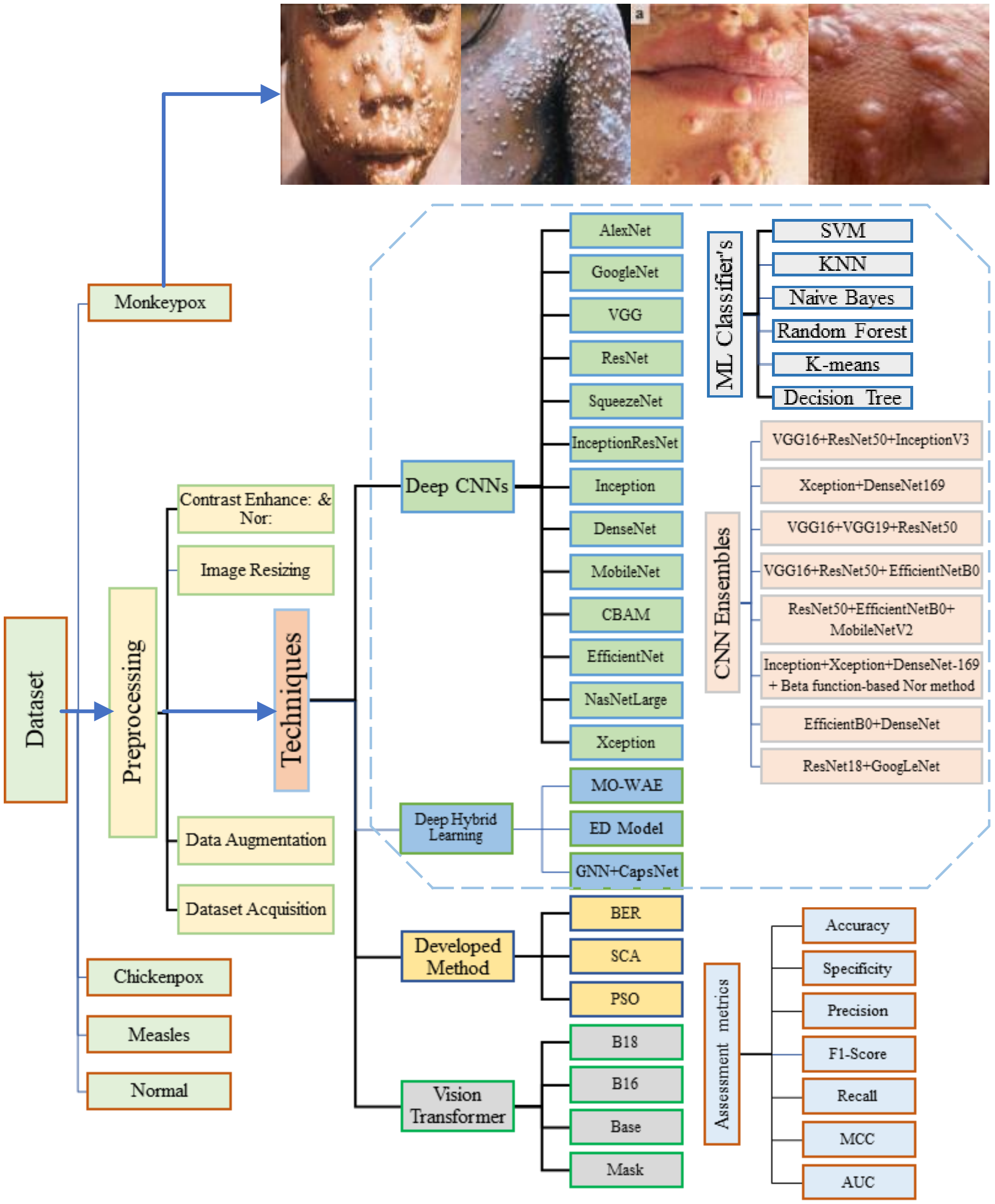

Figure 6: MPox Image Analysis Workflow Utilizing DL Techniques and Performance Metrics.

### 5.1. Deep CNNs

DL-based image detection, particularly with deep CNNs, shows promise in the arena of infectious disease detection, as shown in Table 3, this section presents a performance analysis survey of existing deep CNNs employed for the detection and diagnosis of MPox. The CNN architecture, as illustrated in Figure 7, comprises a sequence of layers, each assigned specific tasks, including convolution and pooling. These layers can be categorized into three types: convolutional, nonlinear, and pooling. Convolutional layers employ distinct filters to the input, resulting in activation maps. Nonlinearities are introduced through subsequent activation layers, and pooling layers reduce the output size from convolutional layers, enhancing robustness and generalization. Fully connected layers extract high-level abstractions and facilitate pixel label-based classification. During the training phase, errors are continuously backpropagated to optimize filter weights and biases. In the following portion, a brief discussion of CNN architecture and the roles of its components. In a convolutional layer, a set of convolutional kernels functions as individual neurons. Each neuron partitions the input image into small receptive fields. The convolution operation can be expressed by equation (1), Here, the symbol $i_c(x, y)$ corresponds to an element found in the input image tensor $l_c$, which is multiplied elementwise by $e_l^k(u, v)$,. This $e_l^k$ represents the index of the kth convolutional kernel in the lth layer. The resulting feature map from the kth convolutional operation can be expressed as $F_l^k = [f_l^k(1,1), \ldots, f_l^k(p, q), \ldots, (f_l^k(P, Q)]$.

Additionally, pooling, or down-sampling, serves to consolidate related information within the vicinity of receptive zones, extracting the most significant response from this local region. Equation (2) illustrates the pooling process, where $Z_l^k$ signifies the aggregated feature-map of the $l^{th}$ layer for the $k^{th}$ input feature map ($F_l^k$), with $g_p$ (.) defining the type of pooling operation. Furthermore, the activation function serves as a component for making choices, enabling the discovery of intricate structures. The activation function for a convolved feature-map is defined in equation (3). In this equation, $F_l^k$ is the output, and it is passed through the activation function $g_a(.)$ introducing non-linearity and yielding a transformed output $T_l^k$ for the $l^{th}$ layer.

To tackle concerns associated with internal variations in feature maps, we utilize Batch Normalization. The transformation of a feature-map $F_l^k$ to a normalized feature-map using batch normalization is expressed in equation 4, where $N_l^k$ denotes the stabilized feature map, $F_l^k$ is the input feature map, and $\mu_B$ and $\sigma_B^2$ signifies the average and standard deviation of a feature map within a small-batch. Furthermore, Dropout is employed to introduce network regularization, enhancing generalization through random deactivating specific neurons or connections having a designated likelihood. The fully connected layer, usually positioned at the network's end, is a worldwide operation that processes input from the previous feature extraction phases and conducts a comprehensive analysis of the output from all preceding layers. This final layer plays a central role in classification tasks. By adopting these components and techniques, our CNN model aims to effectively distinguish MPox from other similar infectious diseases, contributing to accurate disease detection and improved public health outcomes. Numerous advancements and innovative architectural concepts have been incorporated into CNNs. Diverse ideas, such as novel block-based approaches, residual learning, multi-path structures, etc. have been introduced to enhance network depth.

$$f_l^k(p, q) = \sum_c \sum_{x,y} i_c(x, y) \,.\, e_l^k(u, v) \qquad (1)$$

$$Z_l^k = g_p(F_l^k) \qquad (2)$$

$$T_l^k = g_a(F_l^k) \qquad (3)$$

$$N_{l}^{k} = \frac{(F_{l}^{k}) - \mu_B}{\sqrt{\sigma_{l}^{k} + \varepsilon}} \quad (4)$$

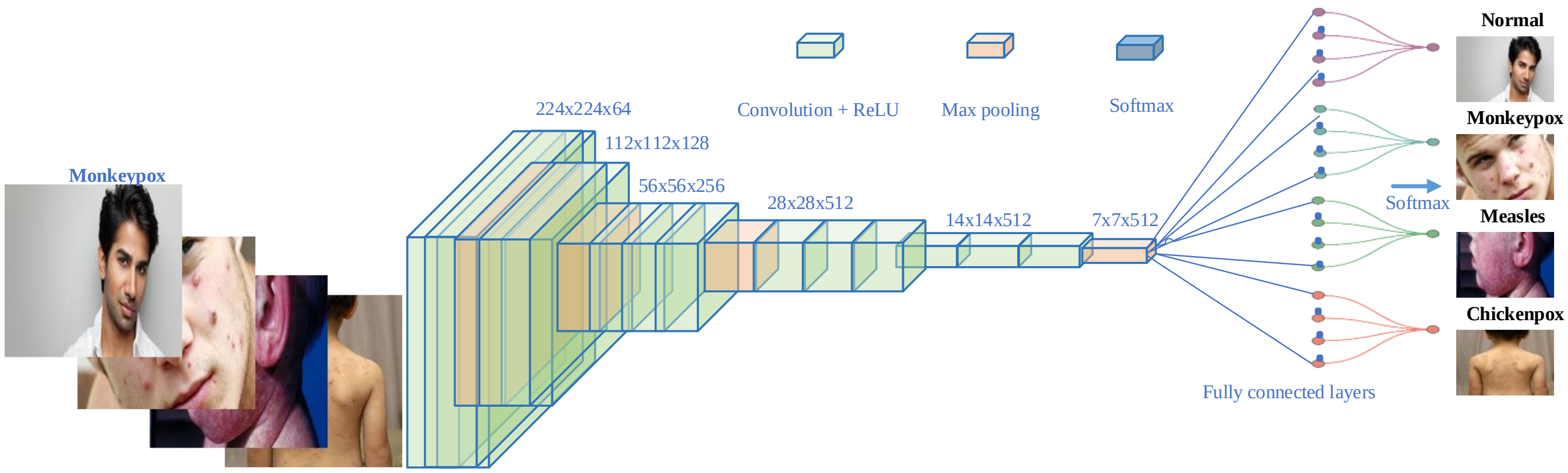

Figure 7: The core CNN architecture for MPox detection.

### 5.1.1. AlexNet

In 2012, AlexNet secured first place at ILSVRC with its eight-layer architecture, featuring a combination of five convoluted layers and three wholly linked layers [43]. The convolutional layers employ kernel sizes ranging from 11x11 to 3x3, operating on 227x227x3 input dimensions. Impressively, AlexNet contains 64 million trainable parameters, and its transition from a sigmoid to a more efficient ReLU activation function resulted in a remarkable decrease in the error rate.

### 5.1.2. VGG/ Spatial Network

In the 2014 ImageNet Challenge, VGG models outperformed, clinching the top positions in both localization and classification categories. Impressively, they achieved the percentage of errors in the test set was 6.8% in classification and 25.3% in localization tracks. This achievement can be attributed to their effective utilization of small filter sizes in deep convolutional networks. VGG16, a prominent CNN architecture created by the Visual Geometry Group, features 16 layers and operates on 224x224-sized images. It employs a sequence of convolutional and pooling layers and, subsequently, a dense layer using softmax activation for image categorization. Notably, this variant contains a more extensive set than VGG-16, in terms of computational layers, resulting in greater memory capacity and processing requirements, specifically featuring 16 convolutional layers [44].

### 5.1.3. Residual Network/ ResNet

ResNet was specifically designed to tackle the problem of vanishing gradients in performance in neural networks as they grow in depth. It accomplishes this by implementing shortcut connections for identity mapping, enabling the creation of even deeper and more trainable networks. An exceptionally deeper ResNet, which comprises eight times as many layers as VGG networks, managed to reduce intricacy, enhance optimization, and achieve a remarkable top-five error rate.

**ResNet-18**, a dependable deep neural network, triumphed in the ILSVRC 2014 Challenge and has showcased its effectiveness in diverse applications, including healthcare. It successfully mitigates the vanishing gradient problem through its shortcut connections. In the ResNet family, ResNet50 and ResNet101 stand out as variations of the ResNet model, primarily differing in their depth. While ResNet50 encompasses 48 convolution layers, ResNet101 includes an additional 3-block layer within the 4th block [45].

**ResNet152V2** represents an altered version of the ResNet, characterized by an extensive set of over a thousand convolutional layers. Distinguishing it from ResNetV1, the version 2 model incorporates batch normalization prior to the weight layer [46], [47].

#### 5.1.4. Google Net

Compared to AlexNet, GoogleNet achieves greater depth and introduces the innovative inception block, leading to its victory in the 2014 ILSVRC. GoogleNet's architecture includes Inception modules that consist of convolutions with 1x1, 3x3, and 5x5 kernel sizes, with 1x1 convolutional layers interleaved to reduce dimensionality within the feature space. Notably, this architecture integrates nine interconnected inception modules [48].

#### 5.1.5. SqueezeNet

SqueezeNet is a compact neural network optimized for visual computing tasks, emphasizing computational efficiency. It incorporates eight Fire modules and strategically employs max-pooling at specific layers. Dropout regularization is implemented to prevent overfitting. Furthermore, adaptations are made to the final convolution layer to suit a two-class dataset [49].

#### 5.1.6. InceptionResNet

Building upon the achievements of Inception and ResNet architectures, researchers integrated the Inception design with residual connections, giving rise to InceptionResNet. The power of the InceptionResNetV2 network is harnessed, a formidable structure encompassing 449 layers, which include Convolution layers, Pooling layers, Batch normalization layers, and more [48].

#### 5.1.7. Inception

The Inception network, pioneered by Google researchers including [48], introduced a novel approach by expanding the network instead of extending its depth. This innovative concept involves employing four convolution layers with different kernel sizes within a specific network depth. This strategy enables the network to capture image features at various scales before forwarding them onto the subsequent layer. Our study utilized an Inception-v3 network with 48 layers.

#### 5.1.8. DenseNet

DenseNet extends the idea of skip connections found in ResNet by introducing dense connections. Within DenseNet, every layer forms a direct connection with all preceding layers, resulting in (L+1)/2 connections for L layers. A key element is the integration of dense blocks, comprising convolution layers with identical feature map sizes but different kernel sizes.

**DenseNet169**, another creation builds on the DenseNet architecture with a focus on optimizing information flow while reducing parameters. It features dense blocks, each composed of multiple convolutional layers interconnected with every other layer in the same block. This interconnected structure efficiently propagates information, enabling the network to learn intricate features and patterns. Additionally, DenseNet169 incorporates transition layers for spatial dimension down sampling, reducing computational demands while introducing regularization.

**DenseNet-121** network, featuring 120 Convolution layers and 4 Average pooling layers. Another variant, DenseNet-201, stands out as a robust CNN architecture designed to enhance efficiency and precision in computer vision applications. Its distinctive feature is the utilization of interconnected blocks, enabling the sharing of characteristics across different layers of the network.

**DenseNet-201** departs from traditional CNNs where each layer takes input only from the preceding layer. Instead, it employs Compact blocks, with every layer in the same block receiving input from the preceding level. This architectural innovation reduces the number of parameters and enhances performance, facilitating more efficient training. The structure of DenseNet-201 comprises dense blocks, transition layers, and global average pooling, culminating with output prediction layers [50].

#### 5.1.9. MobileNet

In 2017, Howard et al. [51] introduced MobileNetV1, a CNN specifically tailored for resource-constrained devices, such as mobile phones. To reduce computational requirements for classification tasks, MobileNetV1 leverages shallow and independent convolution layers. These layers incorporate channel-wise convolution, which uses one filter for every input channel, followed by pointwise convolution that blends the resulting channels from the previous depth-wise convolutional process. The inventive method enables MobileNetV1 to attain superior precision while utilizing considerably Reduced parameters and computational workload in comparison to conventional CNNs.

**MobileNet-V1** incorporates various essential hyperparameters, including the width multiplier, which allows for the adjustment of the channel count in each layer, and the resolution multiplier, which facilitates alterations in the input image size. These adjustments effectively strike a balance between accuracy and computational demands, catering to a variety of devices. Moreover, modifications made to the initial MobileNet design, such as the elimination of the final layer and the addition of new layers, contribute to the refinement of the model, enhancing its data processing capabilities.

Another milestone in the evolution of mobile-friendly CNNs is **MobileNet-V2**, also introduced by Howard et al. in 2017. This CNN architecture is explicitly designed to perform optimally on mobile devices. MobileNet-V2 employs an inverted residual framework featuring residual connections linking bottleneck layers. Like ResNet50, it comprises around 50 convolutional layers, one pooling layer, and one dense layer. This design ensures a more efficient and precise model, ultimately improving the capacity for data processing.

#### 5.1.10. Convolutional Block Attention Module

Convolutional Block Attention Module (CBAM) incorporates two consecutive attention mechanisms: Channel and spatial attention. Following CBAM processing, the output undergoes a pair of dense layers with 256 and 128 units, both utilizing ReLU functions. CBAM serves as an adaptive image refinement module, enhancing critical image features by applying sequential adjustments in the channel and spatial dimensions [52].

#### 5.1.11. EfficientNet

The EfficientNet architecture introduced a systematic scaling approach for CNNs, employing a constant set of scaling coefficients. This architecture comprises three fundamental blocks: stem, body, and final blocks. It's worth noting that the initial and concluding blocks maintain consistency throughout all EN variations, while the core varies. For instance, EN-B0 includes 237 layers, whereas EN-B1 and EN-B2 feature 339 layers, omitting the upper layer. ENB7 is part of the EN model series, denoted 0–7, developed using compound scaling techniques. These models are characterized by three key parameters: alpha, beta, and gamma, with different values for each variant.

**ENB0**, a deep CNN model, employs uniform scaling techniques across all dimensions. It integrates inverted residual blocks from MobileNetV2 and incorporates squeeze-and-excitation techniques within blocks. This model encompasses roughly 237 layers [53] and has demonstrated remarkable performance, boasting an accuracy of about 91.7% on the CIFAR-100 dataset, making it a widely adopted transfer learning (TL) technique among researchers [54].

**ENV2M** represents a better and more computationally efficient version of the original EN. It is structured with seven component sections; each containing its respective modules. EN model variants, such as ENV2L, are built upon a common CNN-based framework. This model employs smaller 3x3 kernel sizes compared to the original EN, reducing memory access cost and the number of parameters. [55].

#### 5.1.12. NasNetLarge

NasNetLarge, a framework developed by Google, employs reinforcement learning strategies to uncover optimal CNN architectures. It aims to determine the most effective parameter settings, including filter size, output channel, stride, and layer count, within the defined search space [56], [57].

### 5.1.13. Xception

The Xception model, inspired by Inception, utilizes separable convolutional layers to drastically reduce trainable parameters. This method has demonstrated its effectiveness across a range of tasks, achieving around 94.5% accuracy on the ImageNet dataset [58], [59]. The details of methods like CNN and other models are provided in Figure 8.

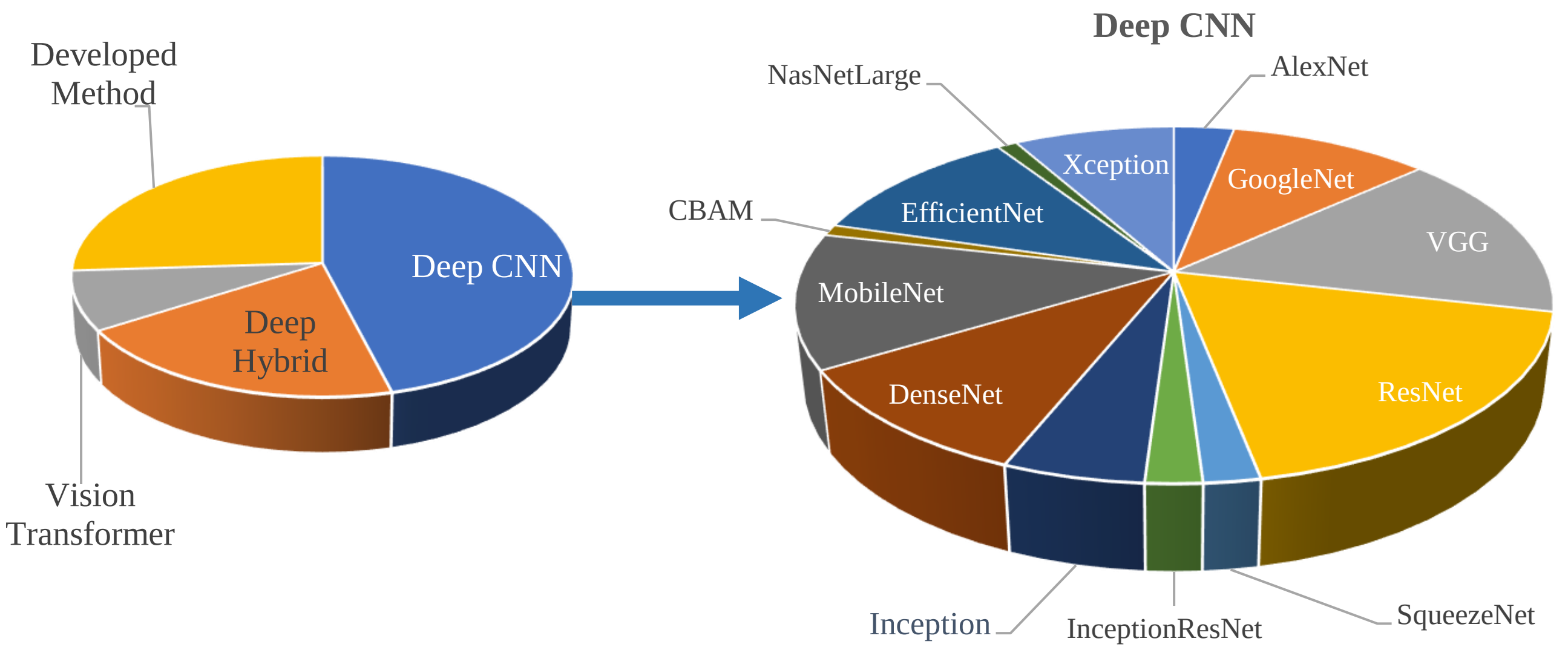

Figure 8: Doughnut Chart: Performance parameters in selected research papers on MPox disease detection Models (CNN, Deep hybrid learning, developed methods, and ViT) up to October 2023.

## 5.2. Deep Hybrid Learning

In the realm of deep CNN training, the risk of overfitting is a well-known challenge. Therefore, the deep hybrid framework focuses on extracting discriminative features that enhance generalization and reduce overfitting as shown in Figure 9. A variety of ML classifiers, like SVM, KNN, Naïve Bayes, Random Forest, and K-means is incorporated, each with unique characteristics tailored to mitigate structural risk. SVM excels at identifying optimal decision boundaries, Naïve Bayes is adept at classification, Random Forest leverages multiple decision trees for classification or regression tasks, KNN handles both classification and regression, and K-means clusters data points based on their similarity. Deep CNNs play a role in reducing empirical risk through strategic hyper-parameter selection. The hybrid approach significantly reduces both training and test errors, ultimately leading to improved generalization. The choice of these classifiers is rooted in their inherent capacity to minimize structural risk, thus contributing to enhanced performance [60].

In deep hybrid learning approach leverages the advantages of deep CNNs and ML classifiers, focusing on extracting crucial features. We obtain High-level features extracted from the top layers of boosted deep CNNs and input them into rival ML classifiers, which encompass SVM, KNN, K-means, and Random Forest. The ML are represented in $f_{SVM}(.)$, $f_{KNN}(.)$, $f_{Random\ Forest}(.)$ as shown in equations (5-7).

$$y_{SVM} = f_{SVM}(\boldsymbol{x}_{DBF}) \quad (5)$$

$$y_{KNN} = f_{Random\ Forest}(\boldsymbol{x}_{KNN}) \quad (6)$$

$$y_{Random\ Forest} = f_{Random\ Forest}(\boldsymbol{x}_{RandomForest}) \quad (7)$$

### 5.2.1 Metaheuristics Optimization Weighted Average Ensemble

A powerful fusion technique, known as the weighted average ensemble, significantly enhances performance. During the training process, foundational models produce probability scores, which are subsequently evaluated for the ground truth classes. Each foundational model is assigned coefficients that are subsequently applied to adjust their predictions. The challenge lies in determining the appropriate weights for each base model. While some researchers experiment with setting ensemble member weights or use grid search, these methods may not consistently yield reliable results due to the difficulty in assessing the contribution of each member. To tackle this issue, the proposed WAE architecture incorporates the Particle Swarm Optimization (PSO) technique. PSO exhibits rapid convergence, efficient exploration of the search space, and eliminates the need for local search. This ensures the optimization of weights in a reliable manner [61].

### 5.2.2 Ensemble Detection Model

The Ensemble detection model is introduced as the novel diagnostic solution during the detection phase. This ensemble approach incorporates three diagnostic methods: Weighted Naive Bayes, DL, and weighted K-NN. These algorithms work on the refined dataset derived from the selection phase, which excludes inappropriate features, to facilitate precise diagnoses of MPox patients. The primary goal of the ensemble model is to amalgamate the results of these three diagnostic techniques using an innovative weighted voting method, thereby enhancing the accuracy of diagnostic outcomes [62].

### 5.2.3 Graph and Capsule-Based Neural Network

Graph Neural Network (GNN) and Capsule-Based Neural Network (CapsNet) emerged as potential alternatives to CNN methods. Nonetheless, concerns have been raised by some researchers regarding their applicability for developing image-based models [63]. These methods continue to undergo active research, while CNN-based approaches remain highly optimized for superior performance with image-based data. Furthermore, it's worth noting that GNN-based techniques often impose higher computational demands, rendering them less efficient than conventional CNN-based methods. They are also constrained in handling a restricted number of data points. Furthermore, GNN-based methods face challenges when dealing with noisy data, which can be a significant hurdle when applied to intricate data points within image-based datasets [64].

### 5.3 Deep CNN Ensembles

Deep CNN ensembles play a crucial role in capturing intricate patterns necessary for distinguishing contrast and texture variations in infected images. This approach enhances both global and local image representations by combining feature maps from prominent deep CNNs [65]. Additionally, the original CNN channels can be merged with auxiliary channels generated through TL, expanding the learning capacity, the performance analysis survey of deep CNNs ensemble and hybrid learning employed for MPox detection is presented in Table 4. A deliberate augmentation strategy, involving a progressive increase in the number of channels, ensures comprehensive and refined learning. In ensemble, a single learner makes decisions by analyzing diverse, image-specific patterns.

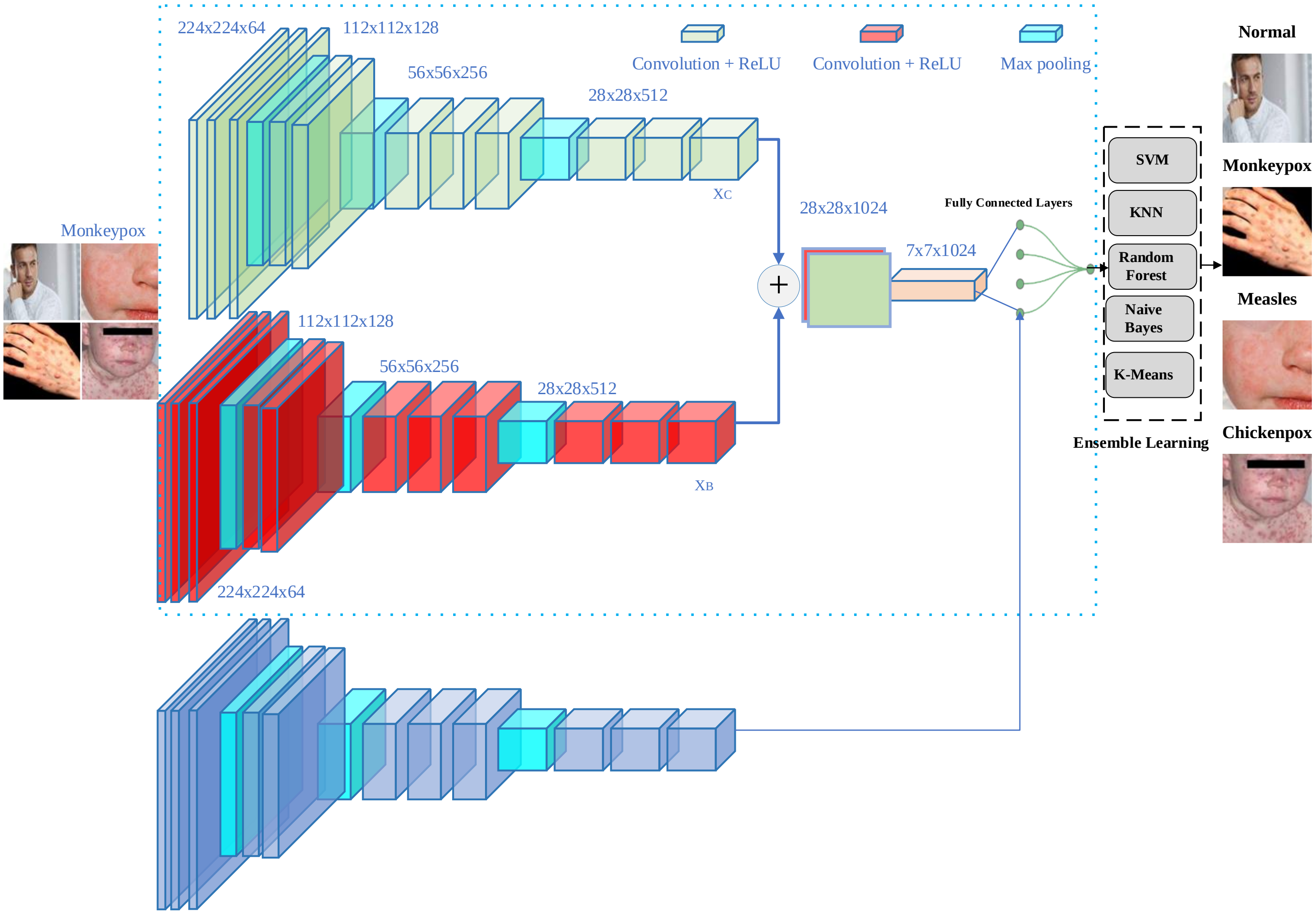

Figure 9: Architecture of CNNs Ensemble with ML Classifier for MPox detection

This augmentation strategy contributes to a thorough and refined learning process, ultimately improving the quality of outcomes. The CNN ensemble architecture is depicted in Figure 10.

$$\boldsymbol{x}_{Boosted=} b(\, \boldsymbol{x}_B || \, \boldsymbol{x}_C) \tag{8}$$

$$\boldsymbol{x}_{DBF} = \sum_a^A \sum_b^B v_a \, \boldsymbol{x}_{Boosted} \tag{9}$$

$$\sigma(\boldsymbol{x}) = \frac{e^{x_i}}{\sum_{i=1}^{c} e^{x_c}} \tag{10}$$

The deep CNNs, referred to as B and C, utilize feature maps represented by $x_B$ and $x_C$, respectively, as depicted in equation (8). Additionally, in some cases, deep CNN channels are concatenated with extra channels generated through TL. The boosting process, denoted as b(.), plays a crucial role in this context. The CNN incorporates densely connected layers with dropout regularization to capture and protect target-specific characteristics while mitigating overtraining [66], [67]. In equation (9), v represents the count of neural units, and equation (10) features the softmax function, denoted as c, which corresponds to the number of classes. This provides a concise overview of the key elements of the model.

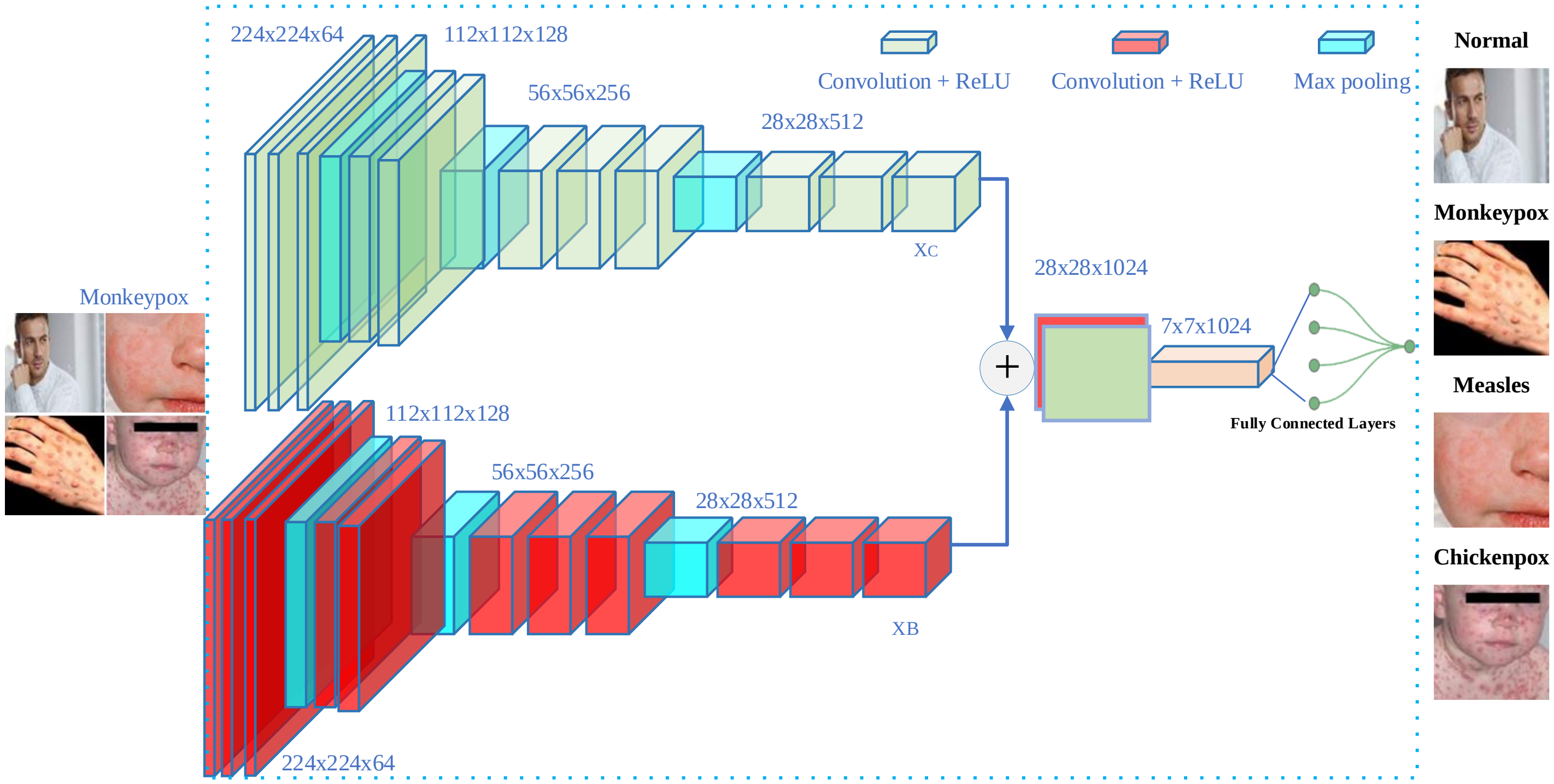

Figure 10: CNN Ensembles Architecture.

## 5.4 Developed Methods

### 5.4.1 BER Algorithm

This BER algorithm calculates Earth's size using Al-Biruni's approach and incorporates optimization techniques to discover the optimal solution while adhering to specified constraints. It employs a population of vectors, each representing an individual member. These vectors undergo exploration and exploitation operations. Exploration seeks promising areas in the search space, while exploitation works on enhancing existing solutions [68].

### 5.4.2 Sine Cosine Algorithm

The algorithm utilizes sinusoidal functions, specifically sines and cosines, to pinpoint optimal solution locations. Random variables are employed to update candidate solutions' positions and velocities, with guidance from current and best solutions to steer the search process. The Sine-cosine algorithm strikes a balance between exploration and exploitation, although its efficiency may decrease in the presence of numerous local optima [68].

### 5.4.3 Particle Swarm Optimization

Particle Swarm Optimization (PSO) mimics the foraging behavior observed in flocking animals, such as birds. It consists of a group of particles, each possessing adjustable values and controlled velocities. These particles track both individual and global extremums and employ these values to adapt their positions and velocities. While PSO exhibits low efficiency and memory usage when compared to other meta-heuristic algorithms, its performance may diminish with a higher prevalence of local optima [68]. Furthermore, the performance analysis Survey of developed DL techniques employed for MPox detection is shown in Table 5.

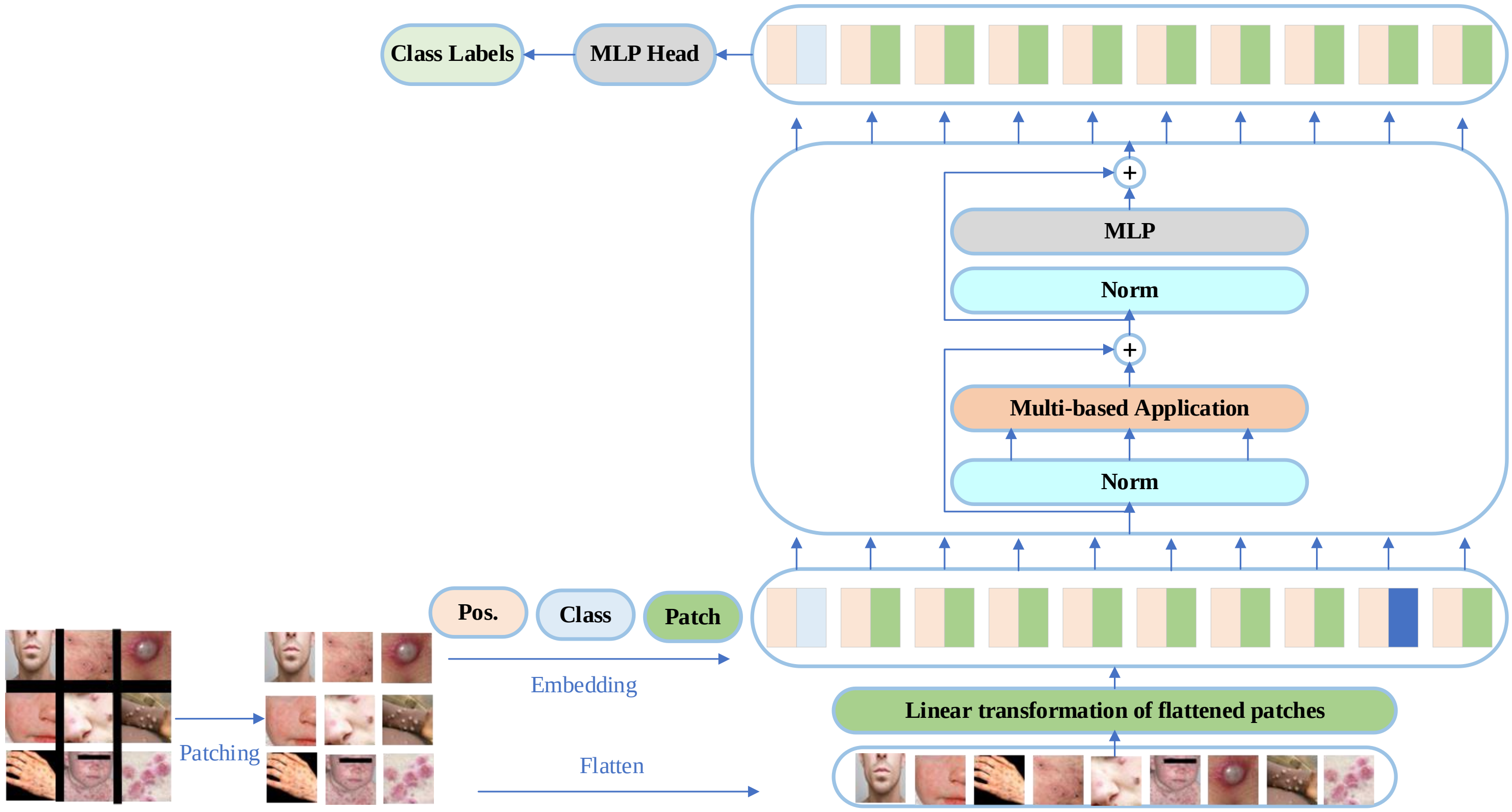

Figure 11: ViT Architecture: Patch-Based Encoding.

## 5.5 Vision Transformer

CNNs, with their limited convolution filter receptive fields, struggle to capture extended dependencies within medical images. In this regard, Transformers have surfaced as a promising substitute for a variety of image-related tasks, including image segmentation, recognition, and object detection. Transformers harness a self-attention mechanism to learn spatial correlations and extended dependencies. To succinctly highlight their differences, transformers, and CNNs each excel in specific areas. CNNs prioritize local contextual information, concentrating on partial image pixels, while transformers shine at capturing both global, long-term dependencies and local representations, Table 6 displays the performance measures of Vision Transformer (ViT) for MPox detection. Figure 11 offers a model overview. In the conventional Transformer architecture, an input consists of a 1D sequence of token embeddings. To address 2D images, the approach involves transforming the image into a sequence of flattened 2D patches, denoted as $X_p$. The image, initially in dimensions $A \times B \times C$, is reshaped into N patches, where $N = \frac{A}{P} x \frac{B}{P}$. In this context, (H & W) denotes the original image's resolution, C stands for the channel count, and P represents the resolution of each image patch. This transformation is integral in treating images as token sequences, a fundamental concept in the ViT architecture [69]. Equation 11 describes the reshaping process, where the parameters are defined as $N = \frac{A}{P} x \frac{B}{P}$. and $D = P * P * C$, with P representing the patch dimensions & C indicating the number of channels. A trainable embedding is incorporated into the sequence of embedded patches, analogous to the [class] token in BERT. During both the pre-training and fine-tuning stages, a classification head is linked to $(Z_L^O)$. The classification head comprises a multi-layer perceptron (MLP) with a single hidden layer in the pre-training stage and a single linear layer during fine-tuning. To maintain positional information, the patch embeddings include straightforward 1D position embeddings, as the adoption of more complex 2D-aware position embeddings did not result in notable performance enhancements. The obtained sequence of embedded vectors serves as the input to the Transformer encoder, which is constructed with alternating layers of multi-headed self-attention and MLP blocks (as described in equations (13) and (14)). Each block is preceded by Layernorm (LN) and includes residual connections. The output state from the Transformer encoder functions as the image representation and is labeled as y equation (15). The MLP itself consists of two layers and employs a GELU non-linearity.

$$X_{Patch}^{NxD} = R(I_{image}^{A\,x\,B\,x\,C}) \tag{11}$$

$$Z_0 = X_{class}; X_p^1E; X_p^2E; \ldots; X_p^NE] + E_{pos,} \quad E \in R^{(p2.c)xD}, E_{pos} \in R^{(N+1)xD} \tag{12}$$

$$Z'l = \mathrm{MSA}(\mathrm{LN}(z_{l-1})) + z_{l-1}, \quad l=1 \ldots L \tag{13}$$

$$Z_l = MLP(LN(z'\,l)) + z'1 \quad l=1 \ldots L \tag{14}$$

$$Y = \mathrm{LN}(z_l^o) \tag{15}$$

Table 3: Performance Analysis Survey of Existing Deep CNNs employed for MPox detection.

| Ref. | Method | Preprocessing | Datasets | Perf. Metrics |
|---|---|---|---|---|
| [70] | Resnet50, EfficientNetV2s | | Erythema Migrans | Acc:(95% CI error) of 86.3% , AUC:0.951 & Kappa of 0.714. |
| [71] | ResNet-18 | DA | MSLD | Acc: 99.8% |
| [72] | MiniGoogLeNet | | MSLD | Acc: 97.1% |
| [73] | VGG-16, LIME | DA & Rs (128 × 128) | MPox 2022 | CI (α = 0.05)<br>**Study1** Dataset Acc% P R Fs Sp<br>Train **97.0** **0.970** **0.970** **0.970** **0.970**<br>Test 83.0 0.880 0.830 0.830 0.660<br>**Study2** Train 88.0 0.860 0.870 0.860 0.890 |
| | | | | Test 78.0 0.750 0.750 0.750 0.830 |
| [74], | VGG-16, ResNet50, InceptionV3, En, MobileNetv2 | DA: & Nor | MSLD | Network P R Fs Acc%<br>VGG16 [6] 0.850 0.810 0.830 81.5<br>ResNet50 [6] 0.870 0.830 0.840 83<br>InceptionV3 [6] 0.740 0.810 0.780 74.1<br>Ensemble [6] 0.840 0.790 0.810 79.3<br>**MobileNetv2 0.900 0.900 0.900 91.1** |
| [75] | ResNet50, EfficientNetB0 | Do, DA & Cv | Lyme disease:1672 | Model Acc% R Sp P Fs AUC<br>ResNet50-141 **84.4** 0.879 0.807 0.831 **0.854** **0.919**<br>EfficientNetB0-187 83.1 0.852 0.809 0.828 0.839 0.909 |
| [76] | VGG19, Xception, DenseNet121, EfficientNetB3, & MobileNetV2 | DA & Rs: (224 x 224 x 3) | MSID | Model Acc% P R Fs<br>VGG19-CBAM- Dense 71.9 0.739 0.729 0.733<br>**Xception-CBAM-Dens 83.9 0.907 0.891 0.901**<br>DenseNet121- CBAM-Dense 78.3 0.805 0.763 0.816<br>MobileNetV2-CBAM-Dense 74.1 0.772 0.722 0.752<br>EfficientNetB3-CBAM-Dense 81.4 0.856 0.811 0.864<br>Ensemble 79.3 0.840 0.790 0.810<br>ShuffleNet 80.0 - - - |
| [77] | ResNet18 | | MSLD | Acc: 99.8%, P: 0.998, R: 0.998, & Fs:0.998 |
| [78] | ResNet-50,DenseNet-121,Inception-V3, MnasNet-A1 SqueezeNet,MobileNet-V2, ShuffleNet-V2. | DA, Cropping, Masking -Rs (224x224x3) | Web-scraping | Methods P R Fs Acc%<br>ResNet50 0.590 0.510 0.550 72.0<br>Inception-V3 0.710 0.530 0.610 71.0<br>DenseNet121 0.710 0.530 0.610 78.0<br>MnasNet-A1 0.630 0.410 0.500 72.0<br>MobileNet-V2 0.730 0.620 0.670 77.0<br>ShuffleNet-V2 0.790 0.580 0.670 **79.0**<br>SqueezeNet 0.550 0.440 0.490 65.0 |
| [68] | AlexNet, VGG-19, GoogLeNet,ResNet50 | Rs & DA | MSID [79] | Acc: 98.8% Sen:0.620 P: 0.998 FS: 0.760 |
| [80] | VGG19, VGG16, Xception, MobileNet & MobileNetV2 using (HHO) | DA & Rs (128×128×3) | MSID + MPID | Dataset Acc% R Sp Fs IoU ROC<br>MSID 97.6 0.952 0.980 0.951 0.909 0.967<br>MPID 97.5 0.948 0.945 0.949 0.905 0.967 |
| [81] | MobileNetV2,VGG16 & VGG19 | DA | MSLD | Model Acc% P R Fs<br>**MobileNetV2 91.4 0.905 0.868 0.883**<br>VGG-16 83.6 0.798 0.745 0.765<br>VGG-19 77.6 0.723 0.718 0.698 |
| [82] | ResNet18, ResNet50, VGG16, Densenet161, EfficientNetB7 & V2, GoogLeNet, MobileNetV2 & V3, ResNeXt-50, ShuffleNet V2, & ConvNeXt models | Rs: (224x224) DA | MSID | Models Acc% R Sp Fs<br>**ResNet-18 98.3 0.966 1 0.983**<br>ResNet-50 96.3 0.931 1 0.964<br>VGG-16 93.0 0.897 0.964 0.929<br>Densenet-161 96.5 0.966 0.964 0.966<br>EfficientNet B7 94.7 1 0.893 0.951<br>EfficientNet V2 96.5 1 0.929 0.967<br>GoogLeNet 96.5 0.966 0.964 0.966<br>**MobileNet V2 98.3 0.966 1 0.983**<br>MobileNet V3 75.4 0.621 0.893 0.720<br>ResNeXt-50 93.0 1 0.857 0.936<br>ShuffleNet V2 79.0 0.655 0.929 0.760<br>ConvNeXt 96.5 1 0.929 0.967 |
| [41] | VGG-16, EfficientNet-B3 | | TBX11K | Acc 88.6% |
| [83] | VGG16, Inception | DA | MPox | **Study 1 Train** Acc% P R Fs Sp<br>VGG16 98.0 0.980 0.980 0.980 0.960<br>**InceptionResNetV2 100 1 1 1 1**<br>ResNet50 56.0 0.320 0.570 0.410 0<br>ResNet101 68.0 0.800 0.680 0.630 0.260 |
| | ResNetV2, ResNet50, ResNet101, MobileNetV2, & VGG19 | | | **MobileNetV2 100 1 1 1 1**<br>VGG19 98.0 0.980 0.980 0.980 0.960<br>**Study 1 Test**<br>VGG16 81.0 0.860 0.810 0.800 0.420<br>InceptionResNetV2 90.0 0.880 0.880 0.770 0.950<br>ResNet50 56.0 0.320 0.560 0.400 0<br>ResNet101 56.0 0.320 0.560 0.400 0<br>MobileNetV2 81.0 0.820 0.810 0.810 0.850<br>VGG19 93.0 0.940 0.940 0.940 0.850<br>**Study 2 Train**<br>VGG16 97.0 0.980 0.970 0.980 0.940<br>**InceptionResNetV2 100 1 1 1 1**<br>ResNet50 72.0 0.510 0.720 0.600 0<br>ResNet101 72.0 0.360 0.500 0.420 0<br>**MobileNetV2 100 1 1 1 1**<br>VGG19 94.0 0.940 0.920 0.930 0.850<br>**Study 2 Test**<br>VGG16 93.0 0.930 0.900 0.910 0.820<br>InceptionResNetV2 98.0 0.980 0.970 0.980 0.950<br>ResNet50 72.0 0.510 0.720 0.600 0<br>ResNet101 72.0 0.360 0.500 0.420 0<br>MobileNetV2 99.0 0.990 0.990 0.990 0.970<br>VGG19 90.0 0.890 0.860 0.870 0.760 |
| [64] | Xception, ResNet-101 | Col, DA, Rs | MPox2022 & MSID | Methods Acc% P R Fs<br>Study 1, Xception 94.0 0.940 0.940 0.940<br>Study2, 80.0 0.800 0.800 0.800<br>**Study 3 ResNet101, 99.0 0.990 0.990 0.990** |
| [84] | Xception, DenseNet | DA, Nor, Rs(24x24) | Data_MPox | Model Acc% P R Fs<br>EfficientNetB3 68.4 0.380 0.720 0.490<br>VGG19 97.8 0.960 0.980 0.970<br>VGG16 96.7 0.950 0.960 0.950<br>ResNet50 70.7 0.340 0.900 0.490<br>**MobileNetV2 98.2 0.990 0.960 0.980** |
| [85] | MiniGoggleNet | | MSLD | Acc: 97.0%, (AUC): 0.760 Loss function: 0.144 |
| [86] | EfficientNetV2s, MobileNetV3, VGG19, ResNet50, DenseNet | DA, Rs (224x224) | MPox Skin Dataset | Methods AUC Acc% R Loss Fs<br>MobileNetV3 Train 0.997 99.1 0.991 0.009 0.978<br>Test (Uniq) 0.990 96.8 0.962 0.034<br>**EfficientNetV2 Train 0.997 99.2 0.992 0.006 0.973**<br>Test (Uniq) 0.989 95.5 0.955 0.034<br>ResNET50 Train 0.992 98.1 0.981 0.015 0.958<br>Test (Uniq) 0.988 93.5 0.935 0.047<br>VGG1 Train 0.981 96.8 0.968 0.027 0.946<br>Test (Uniq) 0.971 92.4 0.924 0.062<br>DenseNet121 Train 0.926 89.6 0.896 0.093 0.895<br>Test (Uniq) 0.927 89.5 0.895 0.096<br>Xception Train 0.851 86.2 0.862 0.140 0.866<br>Test (Uniq) 0.842 85.2 0.852 0.144 |
| [87] | EfficientNet-B4, ResNet-50, MobileNet, Inception-V2 | Nor, DA | MSID & MSLD | **MSID**<br>Technique Acc% Loss Sp R<br>Inception V3 94.0 0.234 0.931 0.952<br>ResNet 50 V2 92.0 0.231 0.897 0.952<br>**MobileNet V2 96.0 0.160 0.931 1**<br>EfficientNet- B4 92.0 0.227 0.897 0.952<br>**MSLD**<br>**Inception V3 93.3 0.246 1 0.880**<br>ResNet 50 V2 88.9 0.272 0.900 0.880<br>MobileNet V2 88.9 0.311 0.900 0.880<br>EfficientNet- B4 88.9 0.277 0.900 0.880 |
| [88] | VGG-19, VGG-16, MobileNet V2, GoogLeNet, & EfficientNet-B0 | Rs (224x224x3), DA | MSLD | **DA**<br>Model Acc% P R Fs<br>VGG-16 97.9 0.981 0.986 0.983<br>VGG-19 98.9 0.984 0.990 0.987<br>EfficientNet-B0 99.2 0.990 0.991 0.990<br>**MobileNet V2 99.3 0.993 0.991 0.992**<br>GoogLeNet 99.1 0.992 0.989 0.991<br>Org<br>VGG-16 71.4 0.715 0.664 0.682<br>**VGG-19 78.8 0.792 0.784 0.787**<br>EfficientNet-B0 75.9 0.7782 0.667 0.707<br>MobileNet V2 71.8 0.703 0.630 0.656<br>GoogLeNet 70.0 0.693 0.624 0.649 |
| [89] | EfficientNetB3, ResNet50, & InceptionV3 | DA, Filtering, Cleaning | MSLD | Models Acc% P R Fs<br>**EfficientNetB3 93.0 0.900** 0.950 **0.930**<br>ResNet50 93.0 0.870 **1** 0.930<br>InceptionV3 73.0 0.670 0.800 0.800 |
| [90] | GoogleNet | Rs, Nor, DA | MSID | Acc: R Sp: *p*-Value N-Value Fs<br>Before FS 85.0 0.940 0.610 0.870 0.790 0.900<br>After FS 87.0 0.950 0.610 0.890 0.790 0.920 |

Table 4: Performance Analysis Survey of Deep CNNs Ensemble and Hybrid Learning employed for MPox detection.

| Ref. | Method | Preprocessing | Datasets | Perf. Metrics |
|---|---|---|---|---|
| | **Deep CNNs Ensemble** | | | |
| [91] | CNN & CNN PA | Rs (299 × 299) | Cancer Img from D, ISIC | **3-way-** Acc (D1:65.6%, D2: 66.0%, CNN: 9.4 CNN – PA: 72.1)<br>**9-way-** Acc (D1: 53.3%, D2: 55.0%, CNN: 48.9, CNN-PA:55.4) |
| [79] | VGG-16 + ResNet50, + InceptionV3 models | DA: & Rs (224 × 224) | MSLD | Network Acc% P R Fs<br>VGG16 81.5 0.850 0.810 0.830<br>**ResNet50 83.0 0.870 0.830 0.840**<br>InceptionV3 74.1 0.740 0.810 0.780<br>Ensemble 79.3 0.840 0.790 0.810 |
| [92] | VGG16, VGG19, ResNet50, ResNet-101, IncepResNetv2, MobileNetV2, InceptionV3, Xception, EfficientB0, EfficientB1, EfficientB2, DenseNet-121, DenseNet-169 En (Xception + DenseNet-169 ) | DA:& Rs (150x150) | MPox 2022 [93] | Methods P R Fs Acc%<br>VGG-16 0.802 0.792 0.790 82.2<br>VGG-19 0.818 0.819 0.810 82.9<br>ResNet-50 0.828 0.829 0.828 84.9<br>ResNet-101 0.827 0.819 0.820 85.0<br>IncepResNetv2 0.839 0.834 0.836 85.4<br>MobileNetV2 0.829 0.812 0.810 84.9<br>InceptionV3 0.8251 0.823 0.822 84.5<br>**Xception 0.850 0.851 0.850 86.5**<br>EfficientNet-B0 0.816 0.813 0.814 84.0<br>EfficientNet-B1 0.837 0.840 0.836 85.1<br>EfficientNet-B2 0.821 0.827 0.821 83.5<br>DenseNet-121 0.831 0.830 0.823 84.2<br>DenseNet-169 0.841 0.837 0.838 86.1 |
| [94] | VGG16, VGG19, ResNet50 | | MPox Skin | Methods Acc% P R Sp Fs<br>VGG16 91.7 0.920 0.920 0.960 0.920<br>VGG19 93.3 0.930 0.930 0.970 0.930 |
| | MobileNetV2, EfficientNet-B0, En1 (VGG-16+ VGG19+ResNet50) , En2 (VGG16+ ResNet50+ EfficientNetB0), En3 (ResNet50 + EfficientNetB0 + MobileNet-V2) | | | ResNet 95.0 0.950 0.950 0.9780 0.950<br>MobileNet 88.3 0.880 0.880 0.940 0.880<br>EfficientNet 90.0 0.900 0.900 0.950 0.900<br>Ensemble1 91.7 0.920 0.920 0.960 0.920<br>Ensemble2 91.7 0.920 0.920 0.960 0.920<br>Ensemble3 98.3 0.980 0.980 0.990 0.980 |
| [95] | Inception + Xception + DenseNet-169 + Beta function-basedNor method | DA, Rs (224 × 224) | MSLD | Acc: 93.4%, P: 0.889, R: 0.968 & Fs: 0.924 |
| [96] | EfficientB0 + DenseNet | DA, Rs (224 × 224) | MSLD | Method / P / R / Acc% / Fs<br>EfficientNet80 / 0.873 / 0.873 / 90.8 / 0.873<br>DenseNet / 0.881 / 0.883 / 88.1 / 0.881<br>Ensembled / 0.946 / 0.945 / 4.6 / 0.945 |
| [97] | ResNet18 + GoogLeNet | | MPox | Acc: 91.57% |
| [98] | CNNs & GWO | DA | MPox PATIENTS Dataset. | / Acc% / P / R / Fs / AUC<br>CNN / 68.8 / 0.706 / 0.879 / 0.783 / 0.615<br>**Run 1 GWO** / **95.3** / **0.957** / **0.982** / **0.969** / **0.927**<br>Run 2 GWO / 93.5 / 0.938 / 0.977 / 0.957 / 0.893<br>Run 3 GWO / 93.3 / 0.934 / 0.979 / 0.956 / 0.887 |
| | **Deep CNN Features and ML Classifiers (Deep Hybrid Learning)** | | | |
| [99] | AlexNet, GoogLeNet, VGG-16, SVM, KNN, nB, DT, RF | DA | MSLD | See table below |
| [100] | (Xception + ResNet-101 + ResNet-50) + DWT + FS + B | Rs:(224x224x3 ), DA | MSID | En / Acc % / R / Sp / P / Fs / MCC<br>**MSID Dataset** |

Results for [99]:

| Bottleneck Features | Classifiers | Class | P | R | Fs | Acc% |
|---|---|---|---|---|---|---|
| AlexNet | SVM | MPox | 0.783 | 0.900 | 0.837 | 84.4 |
| | | Others | 0.909 | 0.800 | 0.851 | |
| | Naïve Bayes | MPox | 0.792 | 0.950 | 0.864 | **86.7** |
| | | Others | 0.952 | 0.800 | 0.870 | |
| | KNN | MPox | 0.667 | 0.800 | 0.727 | 73.3 |
| | | Others | 0.810 | 0.680 | 0.739 | |
| | DT | MPox | 0.650 | 0.650 | 0.650 | 68.9 |
| | | Others | 0.720 | 0.720 | 0.720 | |
| | RF | MPox | 0.810 | 0.850 | 0.829 | 84.4 |
| | | Others | 0.875 | 0.840 | 0.857 | |
| GoogleNet | SVM | MPox | 0.350 | 0.350 | 0.350 | 42.2 |
| | | Others | 0.480 | 0.480 | 0.480 | |
| | Naïve Bayes | MPox | 0.588 | 0.500 | 0.54.1 | **62.2** |
| | | Others | 0.643 | 0.720 | 0.679 | |
| | KNN | MPox | 0.417 | 0.500 | 0.455 | 46.7 |
| | | Others | 0.524 | 0.440 | 0.478 | |
| | DT | MPox | 0.400 | 0.400 | 0.400 | 46.7 |
| | | Others | 0.520 | 0.520 | 0.520 | |
| | RF | MPox | 0.500 | 0.400 | 0.444 | 55.6 |
| | | Others | 0.586 | 0.680 | 0.630 | |
| VGG16Net | SVM | MPox | 0.696 | 0.800 | 0.744 | 75.6 |
| | | Others | 0.818 | 0.720 | 0.766 | |
| | Naïve Bayes | MPox | 0.905 | 0.950 | 0.927 | **91.1** |
| | | Others | 0.957 | 0.917 | 0.936 | |
| | KNN | MPox | 0.842 | 0.800 | 0.821 | 84.4 |
| | | Others | 0.846 | 0.880 | 0.863 | |
| | DT | MPox | 0.783 | 0.900 | 0.837 | 84.4 |
| | | Others | 0.909 | 0.800 | 0.851 | |
| | RF | MPox | 0.833 | 1 | 0.909 | 91.1 |
| | | Others | 1 | 0.840 | 0.913 | |

| | | | | | | | | | | |
|---|---|---|---|---|---|---|---|---|---|---|
| | (Xception + ResNet-101 + ResNet-18) + DWT + FS + B | | MSLD | Bagging | 97.1 | 0.957 | 0.983 | 0.982 | 0.969 | 0.941 |
| | | | | RF | 96.7 | 0.957 | 0.977 | 0.974 | 0.966 | 0.934 |
| | | | | RS | 97.1 | 0.953 | 0.987 | 0.985 | 0.969 | 0.941 |
| | | | | **MSLD Dataset** | | | | | | |
| | | | | **Bagging** | **98.7** | **0.990** | **0.984** | **0.981** | **0.985** | **0.973** |
| | | | | RF | 98.2 | 0.980 | 0.984 | 0.980 | 0.980 | 0.965 |
| | | | | RS | 98.2 | 0.990 | 0.976 | 0.971 | 0.981 | 0.965 |

Table 5: Performance Analysis Survey of developed DL techniques employed for MPox detection.

| Ref. | Method | Preprocessing | Datasets | Perf. Metrics |
|---|---|---|---|---|
| [61] | MO-WAE (DenseNet201, MobileNet, & DenseNet169), (weighted average En technique) | DA, Rs(224x224) | (MSLD) | Acc% 97.8, R 1, P 0.952, Fs 0.976, MCC 0.956 |
| [101] | Maximum entropy algorithm | | Nigeria 116 MPox patients | AUC 0.920 |
| [102] | BER-LSTM | DA | MSLD | Mean absolute error 15.25 |
| [103] | Nine forecasting models | Oversampling | Global MPox Cases | Mean absolute error 146.29 |
| [104] | Multi-layer perceptron, ARIMA model | DA | Website 6 May to 28 July 2022 | Mean absolute error 32.59 |
| [105] | BERT-based ML model. | | 170 facts & 55 wrong (Internet) | Acc: 96% |
| [106] | Hybrid CNN–LSTM | | MPox 61379 tweets | Acc: 94% |
| [107] | BERT, BERTopic | | 352,182 Twitter posts | ----------- |
| [108] | Latent Dirichlet allocation | | 15,936 Twitter posts from Germany | ----------- |
| [109] | BiLSTM | DA | DNA sequences of MPox & human papillomavirus [69] | Acc: 96.1% Fs:0.988 |
| [110] | customized CNN + BERSFS | Scaled & Nor of 227 × 227 | MSID #Img = 770 | Acc% 98.8, R 0.857, Sp 0.992, pValue 0.760, nValue 0.996, Fs 0.805 |
| [111] | Modified DenseNet-201 | FS, Rs, DA | MSID | Metrics / Org: dataset / DA dataset<br>P 0.932 / 0.989<br>R 0.932 / 0.989<br>Fs 0.932 / 0.989<br>Test acc%: 93.2 / 98.9<br>AUC 0.992 / 0.9997 |
| [112] | Customized CNN | DA, Rs | MSLD | P:0.980 R:1 Fs:0.990 TP:49 FP:1 FN:0 TN:50 Acc:99.0% |

Table 6: Performance Analysis Survey of ViT employed for MPox detection.

<table>
<tr><th>Ref.</th><th>Method</th><th>Pre-processing</th><th>Datasets</th><th>Perf. Metrics</th></tr>
<tr><td>[113]</td><td>GoogLeNet,Places365, GoogleNet,SqueezeNet, AlexNet, ResNet-18, & ViT B18</td><td>DA, Rs (224x224)</td><td>MSLD [79]</td><td><table>
<tr><td>Method</td><td>Acc %</td><td>R</td><td>Sp</td><td>P</td><td>Fs</td></tr>
<tr><td>ViTB18</td><td>71.6</td><td>0.793</td><td>0.685</td><td>0.498</td><td>0.611</td></tr>
<tr><td>Proposed</td><td>99.5</td><td>0.994</td><td>1</td><td>1</td><td>0.995</td></tr>
</table></td></tr>
<tr><td>[114]</td><td>ViT</td><td></td><td>MPox: 1300 Img</td><td>Acc: 93.0%</td></tr>
<tr><td>[115]</td><td>ViT</td><td>DA</td><td>MSLD</td><td>P:95.00 R:95.00 Fs:95.00 Acc:94.69%</td></tr>
<tr><td>[114]</td><td>ViT base</td><td>DA, Acquisition</td><td>MSLD</td><td><table>
<tr><td>Model</td><td>Acc%</td><td>P</td><td>R</td><td>Fs</td></tr>
<tr><td>SVM</td><td>65.0</td><td>0.690</td><td>0.780</td><td>0.730</td></tr>
<tr><td>K-NN</td><td>84.0</td><td>0.820</td><td>0.810</td><td>0.820</td></tr>
<tr><td>RestNet50 with TL</td><td>91.0</td><td>0.910</td><td>0.900</td><td>0.900</td></tr>
<tr><td>ViT</td><td>93.0</td><td>0.930</td><td>0.910</td><td>0.920</td></tr>
</table></td></tr>
</table>

# 6 Evaluation Setup:

## 6.1 Cross Validation and Performance Metrics:

Various cross-validation techniques were employed in the experiments, including k-fold and holdout. Given the limited size of the collected dataset, adapting the cross-validation concept was crucial to enhance the dataset's effectiveness. The subsequent steps involved applying various ML classifiers to categorize image classes. Typically, the cross-validation process partitions the initial images into the test and train at a 20:80 ratio. Aug was used on the train and validation sets, while the test set exclusively contained the initial images. To evaluate the classification models' performance, a comprehensive set of metrics was employed. These metrics, calculated using standard equations, assess the model's effectiveness in distinguishing MPox-infected individuals from non-MPox cases. They involve parameters like False Positives (FP), True Positives (TP), False Negatives (FN), and True Negatives (TN). The assessment metrics include accuracy, precision, F1-score, recall, MCC, sensitivity, and AUC, as detailed in

Table 7 and Figure 12.

## 6.2 Hyperparameter Settings:

This study utilizes the MPox dataset, which records the clinical characteristics of human MPox infection during a recent outbreak. The dataset comprises 11 attributes and a target variable that indicates the existence or non-existence of MPox. The selection of batch size, epoch count, and learning rate significantly impacts the model's performance. To identify the optimal hyperparameters through experimentation, we employed grid search. This approach involved exploring various combinations of learning rates, batch sizes, Regularization, and other parameters to achieve the highest accuracy and lowest loss during model training. Regularization serves as an adjunct technique with the primary objective of enhancing a model's generalization capabilities, ultimately leading to improved performance on the test dataset. This approach encompasses a range of factors, including the characteristics of the loss function, the optimization method used for minimizing the loss, and the incorporation of various other techniques.

Table 7: Common Performance Measures.

| Metrics | Abb: | Metrics | Description |
|---|---|---|---|
| Accuracy | Acc | $\frac{TP + TN.}{TP + TN + FP + FN}$ | Quantifies the fraction of accurately labeled instances out of all instances considered. |
| Specificity | Sp | $\frac{TN}{TN + FP}$ | Evaluate the model's capacity to accurately recognize TN. |
| Precision | P | $\frac{TP}{TP + FP}$ | Quantifies the ratio of TP instances correctly recognized among all instances categorized as positive. |
| Recall | R | $\frac{TP}{TP + FN}$ | Measures the model's capability to detect all positive instances. |
| F1-Score | FS | $\frac{2*(P*R)}{P+R}$ | The F1-score, appropriate for imbalanced class distributions, is the average of precision & recall. |
| Matthew's Correlation Coefficient. | MCC | $\frac{(TP \times TN - FP \times FN)}{\sqrt{((TP+FP)x(TP+FN)x(TN+FP)x(TN+FN))}}$ | It considers all four metrics (TN, FN, TP, FP) to assess classification quality. |

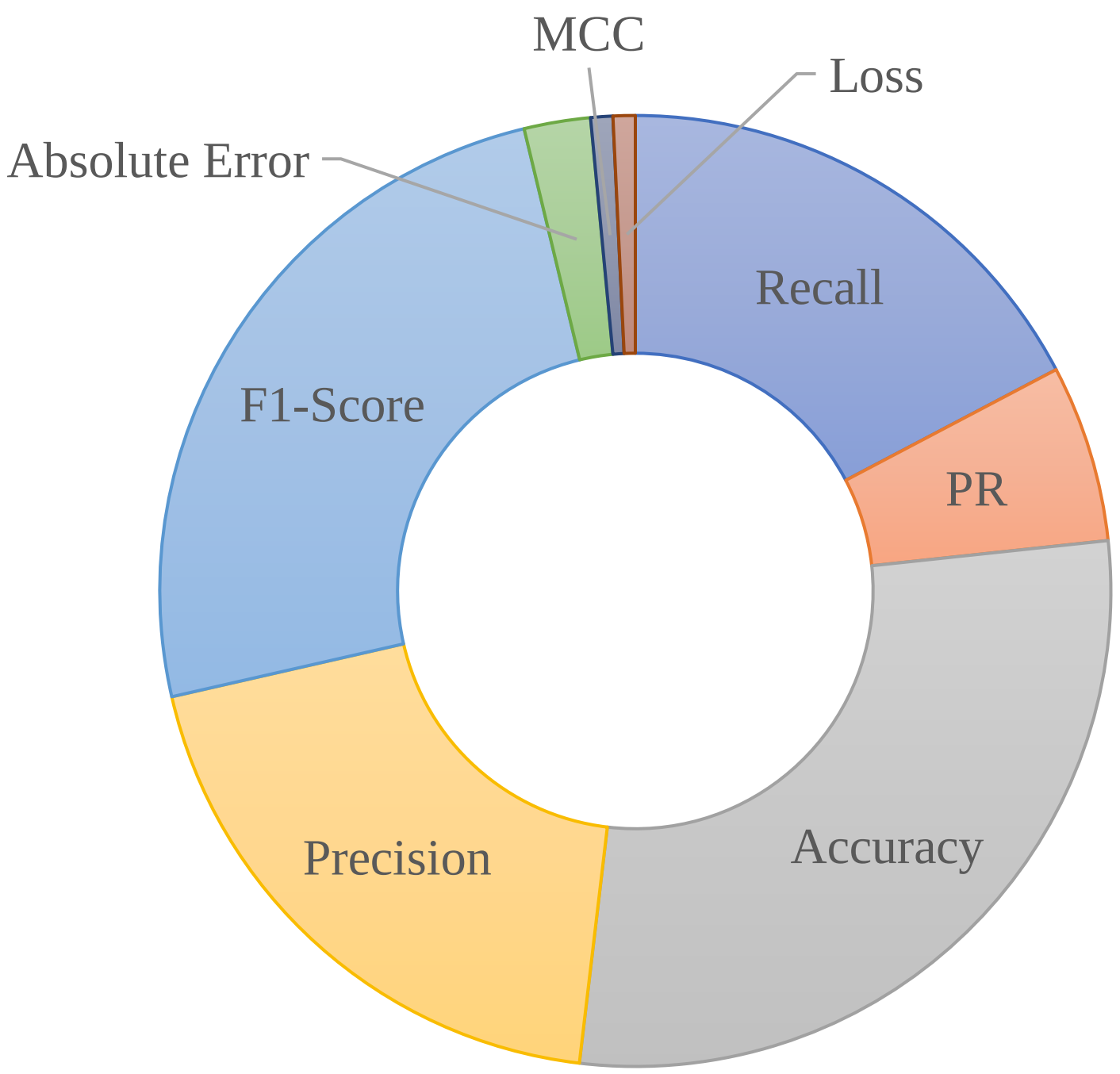

Figure 12: A Doughnut Graph displaying performance parameters from selected research papers on MPox disease detection up to 2023, including metrics like Accuracy, Precision, F1-Score, Recall, PR, and Loss.

### 6.3 Feature Selection and Extraction

Optimizing prediction model performance hinges on effective feature selection. To identify the most influential features impacting MPox, a correlation analysis was conducted. This analysis enables to pinpointing of variables strongly correlated with the target variable, facilitating the choosing of the most pertinent features while mitigating approach overfitting. The correlation matrix, depicted as a heatmap, visually portrayed the interrelationships between feature pairs. In the realm of CNN architecture, feature extraction plays a pivotal role. It entails dimensionality reduction to bolster the efficiency of DL models. CNN models are characterized by multiple layers meticulously crafted to recognize and retrieve relevant attributes from the data.

## 7 Challenges

In recent years, numerous DL methods have emerged for diagnosing MPox. Most of these approaches employ CNNs for MPox screening and analysis. However, these methods commonly encounter certain challenges below, as detailed in Figure 13.

### 7.1 Lack of Standard Dataset

CNNs rely heavily on extensive annotated datasets to produce accurate results, given their numerous optimization parameters. However, due to MPox being a relatively new disease, images are scarce in standardized formats suitable for training and evaluating CNN models. This shortage of information stems from multiple factors, like the limited availability of data, resource constraints in emergency settings, and ethical concerns. The challenge extends further when considering the need for a significant amount of annotated data, especially for deep CNNs in a supervised setting. Properly formatted information labeling is essential for training models, particularly for segmentation tasks, which demand a substantial annotated dataset. The scarcity of sufficient data significantly hampers the training of DL models.

### 7.2 Human Subjectivity

During public health emergencies, particularly in epidemics and pandemics, hospitals and radiologists face a substantial burden that impacts their performance. The scarcity of experienced radiologists is evident in various medical domains. Additionally, when analyzing medical images, radiologists tend to exhibit greater specificity and lower sensitivity in identifying infectious diseases.

### 7.3 Computational Limitations

DL models demand substantial resources, necessitating significant computational power and storage capacity. Attempting to simulate DL models on a CPU is notably slow, time-consuming, and ultimately impractical. Efficient handling of the heavy processing demands is achievable through GPU-based systems or high-power processing clusters. Deep CNNs come with substantial computational demands, necessitating ample computing resources for training. Regrettably, the scarcity of these resources poses a significant hindrance to the advancement of DL models.

### 7.4 Optimization

The performance of DL methods can vary depending on hyperparameters, including model depth, optimizer choice, loss function, and image preprocessing. Several researchers have used various deep CNN architectures without implementing a robust hyperparameter optimization strategy for MPox detection, leading to minimal discrepancies in the outcomes across different architectures.

### 7.5 Model Assessment Strategies

In the quest for a robust deep CNN model within diagnostic systems, the challenge lies in achieving generalization. Clinical deployment requires thorough validation, but many published papers use a single dataset for training, validation, and testing. Regrettably, such methods often exhibit limited transferability when applied to datasets from diverse origins. In a pandemic scenario, time constraints often prevent thorough testing, leading to incomplete results. Furthermore, many researchers have sourced information from various online resources, potentially introducing repeated images. This duplication can lead to overfitting and unreliable results since the identical image might appear in both training and evaluation sets. Detecting and addressing data duplication becomes vital to prevent overuse. Additionally, comparing different techniques becomes challenging due to disparities in data distribution and sample allocation.

### 7.6 Models Generalization

Achieving strong generalization and peak performance in a clinical context often requires collecting data from diverse sources. However, gathering data from various domains presents challenges due to privacy concerns and complexities in data integration. Every dataset showcases distinct data gathering features shaped by the tools and clinical procedure applied. Consequently, integrating data from various sources requires careful consideration of hardware specifications for capturing images, visual clarity, and image file format. Compounded by the restricted volume of information and the inadequate coverage of various MPox variations, many models exhibit reduced effectiveness when applied to external testing. Similarly, models trained on demographic regions often show limited generalization when used in regions beyond their training scope.

### 7.7 DL Clinical Evaluation

AI-driven diagnostic systems encompass stages such as model creation, deployment, and updates. Nonetheless, the conversion of these diagnostic systems into experimental setups tends to progress at a sluggish pace. Much of the effort has been directed toward developing new models, often overlooking the critical deployment phase. This delay is primarily attributable to restricted access to experimental data, interoperability challenges within scientific environments, and authorized and fair considerations, which collectively hinder the translation process.

### 7.8 Interoperability Challenges

DL-driven diagnostic systems are often crafted on various platforms using a range of frameworks and libraries. Deploying these DL systems in hospital settings necessitates the establishment of standard operating procedures. These steps are essential to ensure smooth compatibility between hospital equipment and diagnostic systems. Achieving this integration requires transforming both DL models and radiographic images into specialized formats, in addition to optimizing for hardware-specific requirements.

### 7.9 Pandemics Expertise/Facilities

Recent pandemics like COVID-19 presented substantial global challenges, emphasizing the importance of readiness for potential future crises. This holds, especially considering the increasing incidence of MPox cases outside endemic regions. Timely and accurate detection is a linchpin for infection control. However, misdiagnosis and variations in clinical presentations across regions can complicate this process. Should another pandemic like MPox emerge, low-resource countries, lacking governance expertise, funding, robust recording and tracking systems, healthcare facilities, trained personnel, fundamental diagnostic equipment, and lab facilities, will encounter severe difficulties. Notably, a significant portion of expenses incurred during a viral pandemic is directed toward healthcare costs, encompassing hospital resource utilization, medical supplies, pharmaceutical expenses, and quarantine-related costs.

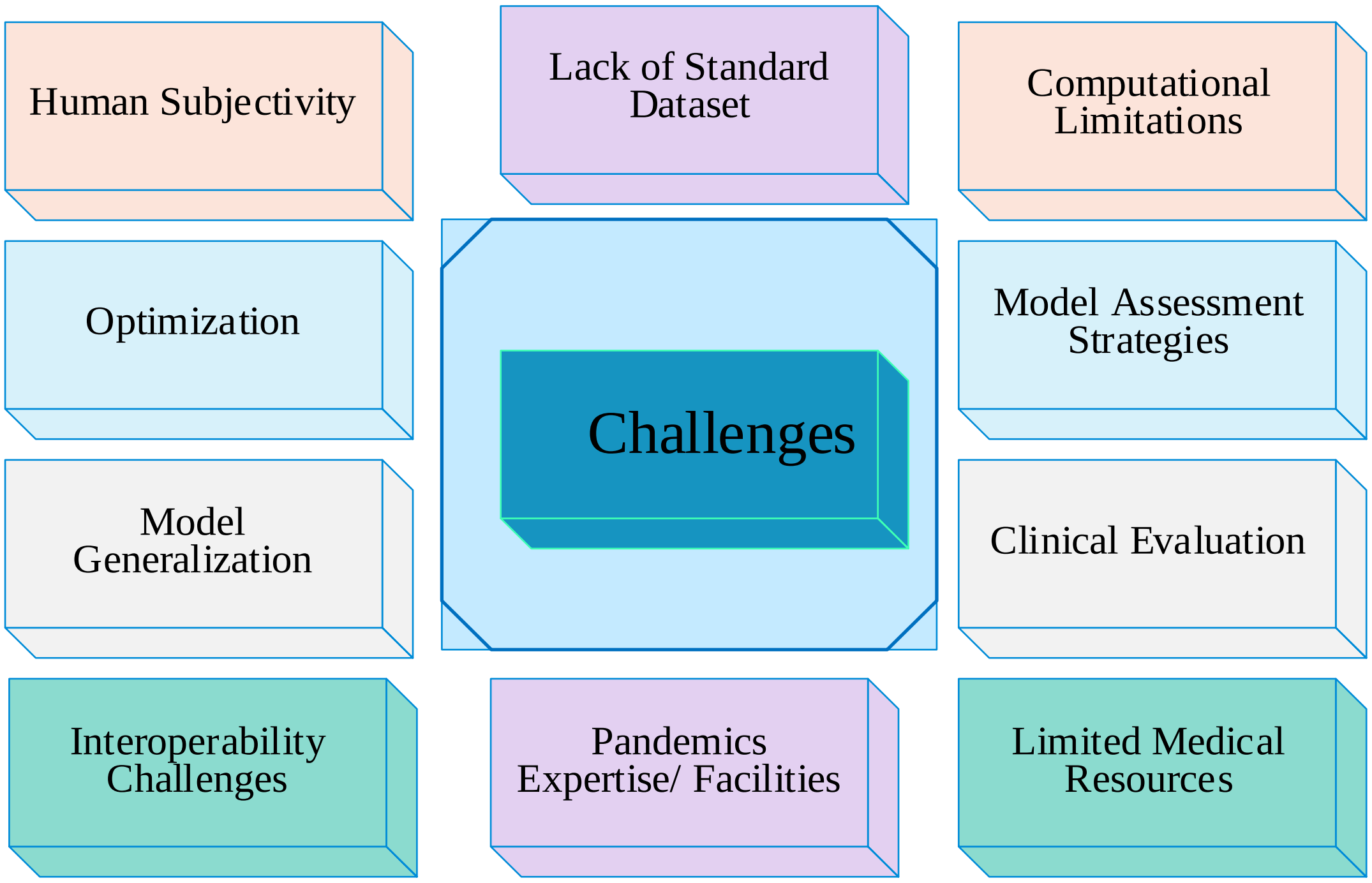

Figure 13: Challenges in MPox Detection.

### 7.10 Limited Medical Resources

Non-compliance with medical treatment and vaccination can be attributed to cultural variations across regions, social prejudice, gossip, and fear. Healthcare workers in developing nations may be without the necessary expertise and confidence to identify medical characteristics indicative of a bleak forecast, and they may not have access to proper safety precautions and equipment during an outbreak. The mental health impacts observed in patients, healthcare workers, and communities during the pandemic may also pose challenges in future outbreaks.

## 8 Limitations

### 8.1 Data Accessibility and Quality

Typically, medical datasets consist of a limited number of publicly accessible labeled medical images accompanied. The scarcity of labeled data arises from concerns related to patient privacy and the constrained time available for labeling, particularly during the pandemic. One significant limitation in the context of MPox is the scarcity of publicly available data, suitable for DL model training. Freely available datasets are frequently limited in size, constraining the application of information-intensive DL methods. DL solutions require substantial information for unbiased, non-overfitting models. Encouraging and facilitating MPox data collection and sharing is crucial for meaningful performance comparisons.

### 8.2 Data Filtration

It is essential to eliminate erroneous or misleading MPox information to ensure the accuracy of input data in AI models. Thirdly, distributing data, particularly delicate unprocessed data, presents security risks that require protection against potential misuse or cyberattacks. Data structuring is challenging, particularly given the abundance of unstructured online data, making it difficult to extract meaningful insights.

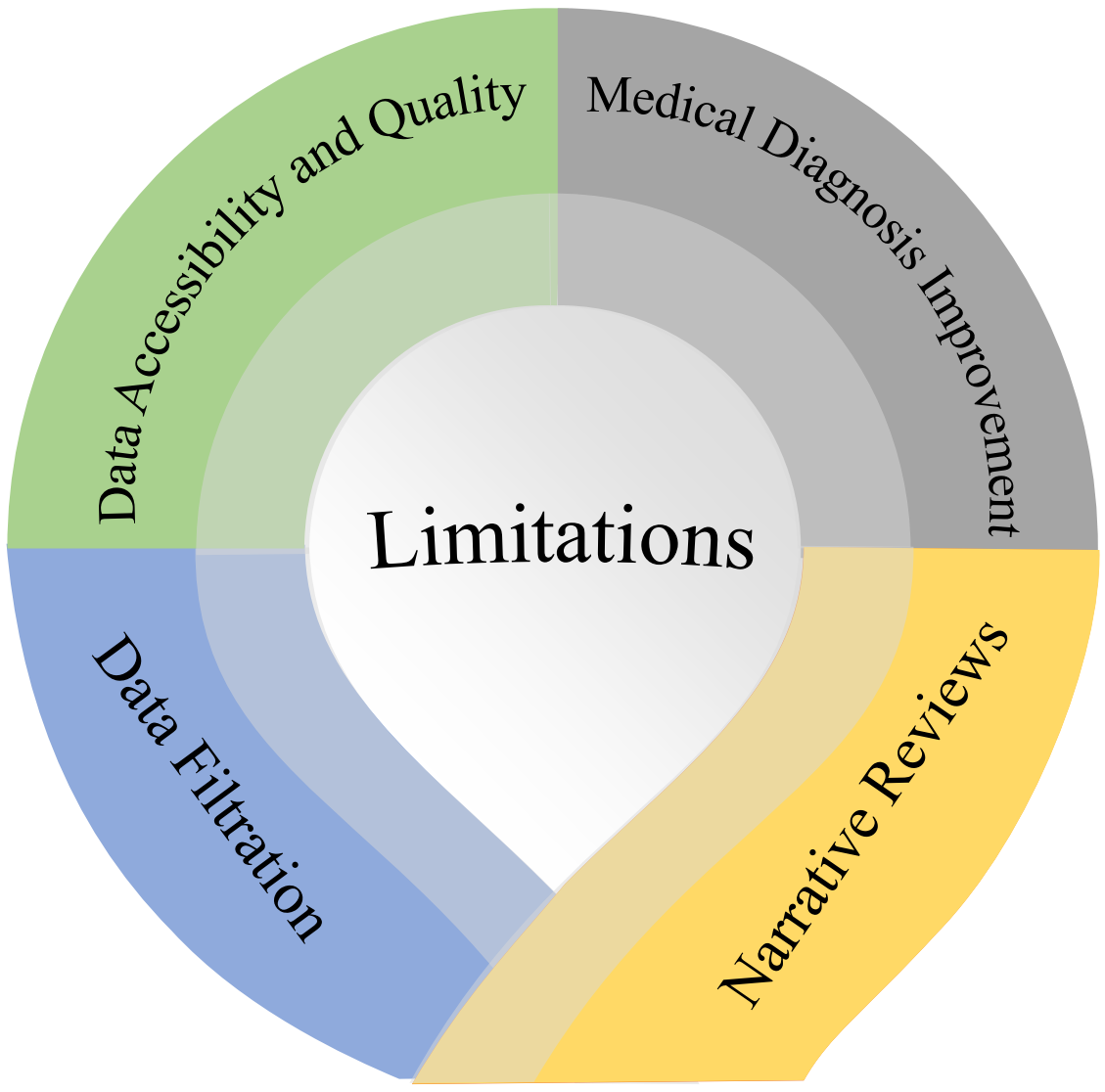

Figure 14: Limitations of MPox Detection

### 8.3 Medical Diagnosis Improvement

DL algorithms are susceptible to incorrect results and errors. Combining DL with other ML approaches, such as self-guided and reinforcement learning, could enhance overall diagnostic accuracy. While many studies rely heavily on skin lesion images for MPox diagnosis, diversifying diagnostic approaches, including blood tests and laboratory markers, can enhance result reliability. Additionally, testing models on datasets covering various virus mutations is crucial.

### 8.4 Narrative Reviews

It's important to highlight that despite the value of comprehensive narrative reviews in improving the detection and understanding of disease dynamics, most researchers often do not conduct such in-depth reviews, which can be considered a limitation in harnessing this resource. A comprehensive narrative review serves as a valuable resource for information on viral diseases over the past three decades. It offers authentic and reliable data from multiple verified sources, providing current data regarding recent viral epidemics. This review enables comparisons between the MPox Virus and the clinical features, signs, and symptoms of various viral illnesses to improve diagnoses and aid in understanding disease spread, treatment, and prevention. The limitation is shown in Figure 14.

## 9 Applications

### 9.1 DL Application for Treatment in MPox

In therapeutics, DL plays a significant involvement in identifying pathogens, image-driven diagnoses, and model-supported diagnostics. It aids healthcare providers and researchers in understanding the mechanisms behind disease causation and interactions between hosts and pathogens [19]. ML algorithms have demonstrated their effectiveness in pharmaceuticals and healthcare, especially in fields like radiology, the analysis of large datasets, early disease identification, and individualized healthcare. Unstructured electronic medical records are efficiently categorized using DL models. DL has given rise to various classification models (MIDDM) that aid in infectious disease detection. Bayesian models, decision trees, and logistic regression are employed to assess the predictive precision of infectious disease outcomes. The MIDDM model significantly enhances disease detection precision [28]. DL harnesses the power of deep neural networks to classify extensive datasets comprising images, text, and audio, thus advancing our understanding of disease severity and type [116]. As the volume of data grows, the advantages of DL networks become even more pronounced. The application, including DL applications for treatment in MPox, other applications of AI in combating MPox, and web-based information on MPox, is shown in Figure 15.

### 9.2 Other Applications of DLin Combating MPox

MPoxV, part of the Orthopox genus within the Poxviridae clan, is responsible for human MPox, a zoonotic ailment:

#### 9.2.1 Transmission and Pathogenesis

MPoxV has an interval of incubation ranging from 5 to 21 days. It initially enters the host through airway epithelial cells, as well as fibroblasts, keratinocytes, and skin endothelial cells. Within the skin, it primarily targets keratinocytes, leading to distinctive changes observed during the vesicular and pustular stages. The virus then spreads to draining lymph nodes, leading to the development of primary viremia, and then spreading to organs such as the spleen and liver, which results in secondary viremia. Notably, MPoxV finds various animal hosts, including gnawing mammals, murids, rats, and primates, and can be transferred through physical contact with animals, involving blood, bodily fluids, airborne particles, and contact with infected sores. Furthermore, transmission from one person to another is feasible, transmitted via proximity and respiratory droplets. There is also the potential for vertical transmission from mother to fetus or in childbirth, significantly increasing the risks of miscarriages and perinatal death.

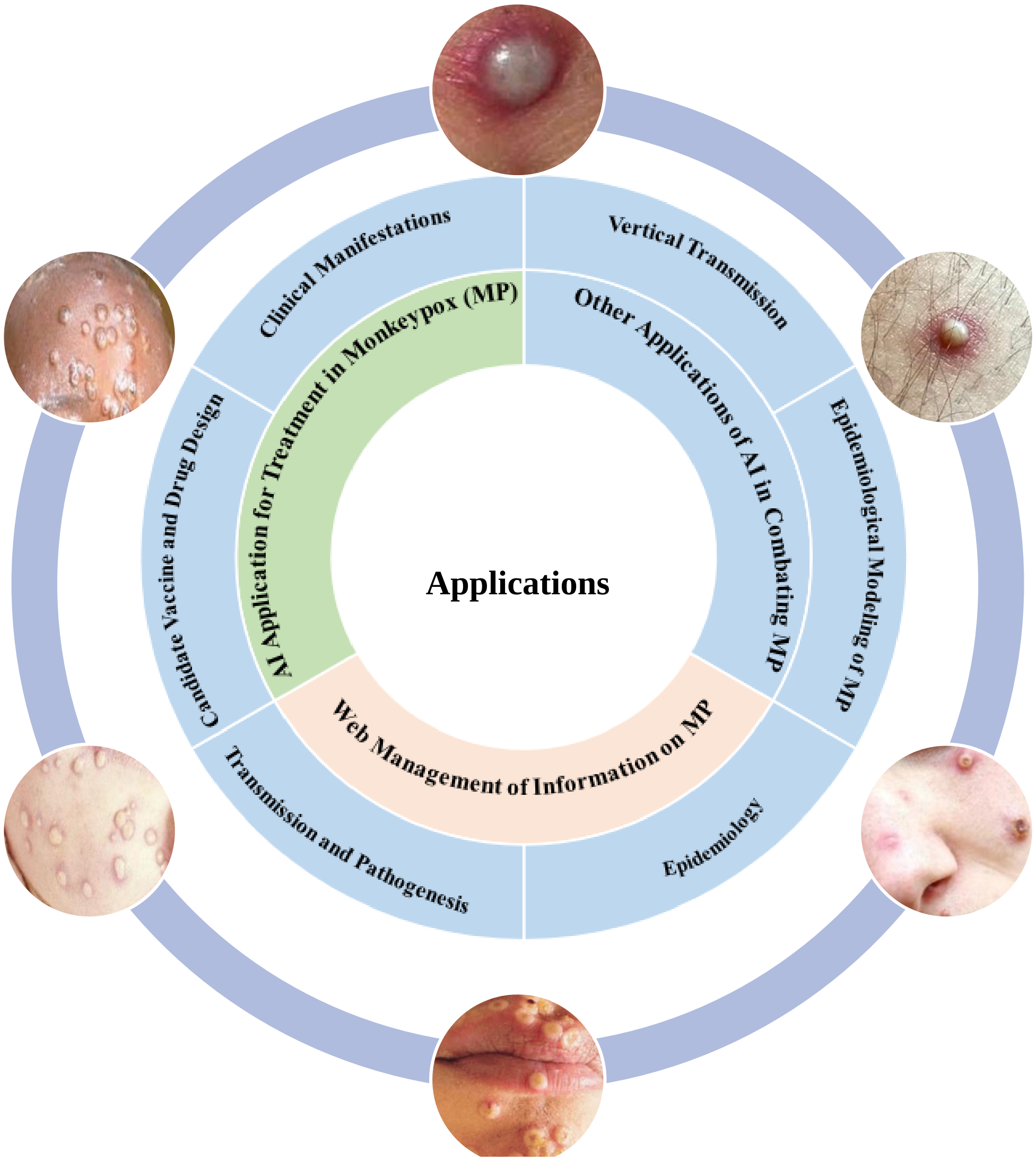

Figure 15. Application for MPox detection and treatment.

### 9.2.2 Epidemiological Aspect

MPox is predominantly prevalent in countries across sub-Saharan Africa but shows a slow expansion into non-endemic regions. The virus is classified into two subgenera, one associated with Western African regions and Congolese regions, the latter is associated with Intense Surges and Elevated fatality rates. The initial MPox outbreak beyond African borders took place in the US in 2003, primarily associated with the West African clade. Reported cases in the UK, Israel, USA, and Singapore have often been connected to travel from Nigeria. Recent data from the WHO underscores a notable increase in cases in 2022, with the majority concentrated in the WHO European Region.

MPoxV infections seem to exhibit a higher prevalence among males, notably within the gay and bisexual population. Alarmingly, mortality rates are more elevated among younger children than adults. While the precise transmission routes from endemic to non-endemic areas remain unclear and travel-related evidence is limited, existing data strongly advocate for the implementation of preventive measures. Further research into the pathophysiology of transmission is imperative to refine prevention strategies, detection, and treatment protocols.

### 9.2.3 Epidemiological Modeling of MPox Infection Spread

AI models have proven highly effective in forecasting MPox outbreaks [101]. harnessed a maximum entropy algorithm, utilizing data from 116 spatially distinct prior MPox infection cases in Nigeria spanning from 2017 to 2021. This model, with top contributing factors such as rainfall, population density, altitude, and low and high temperatures, provided accurate predictions of conditions conducive to MPox spread and the geographical regions at risk, thereby facilitating resource allocation. A polynomial network trained on MPox case information gathered from May 6$^{th}$, 2022, to July 28$^{th}$,2022, to create a prognostic model with the ability to forecast MPox cases for the upcoming 100 days [117]. The "BER-LSTM" method, which employs a long short-term memory network and is optimized employing the AI-Biruni Earth Radius algorithm, was introduced for predicting MPox disease transmission [61]. Through the incorporation of statistical methods, it achieved a mean bias error of only 0.06%. A predictive model utilizing regression analysis to forecast MPox outbreaks [60]. This model outperformed eight other algorithms [60]. In the context of MPox case forecasting, [61] conducted a comparative analysis of different time-based DL models. Among these models, the multilayer perceptron emerged as the most dependable choice.

### 9.2.4 Development of Vaccines and Drugs

Multiepitopic vaccines are gaining recognition for their ability to stimulate specific immune responses by targeting conserved antigenic regions in antigenic sequences, reducing undesired reactions. In this study, researchers retrieved MPox proteins from the Virus Pathogen Analysis Resource (ViPR) Database [62]. They identified nine overlapping antigenic regions and formulated multi-antigenic vaccines, integrating effective Adjustments aimed at strengthening immune reactions. [63]. In a separate study [118], 176 genomic-encoded protein structures were assessed as potential MPox vaccine candidates using immunoinformatic analysis. This screening process resulted in a final model with an impressive affinity of 98.4 kcal/mol to the MPoxV. Additionally, stimulating B and T lymphocyte immune responses is focused on three extracellular antigenic proteins, providing valuable insights into potential vaccine creation against MPox [119].

### 9.2.5 Clinical Manifestations

MPox patients commonly exhibit symptoms including chills, fever, myalgia and back discomfort, tiredness, lymph node swelling, and cutaneous eruptions. These rashes are primarily observed on the extremities, chest, genitalia, or anus. The disease progresses through stages, with the initial phase characterized by fever, severe headache, and lymphadenopathy. Notably, lymphadenopathy is a key distinguishing feature of MPox from other ailments. Rashes also develop at various sites, and approximately 95% of patients develop facial rashes. These rashes may change in appearance over time, eventually drying up and falling off. It's worth mentioning that severe cases are more commonly seen in children. Generally, clinical manifestations resolve without the need for treatment. However, complications, including encephalitis, bronchopneumonia, sepsis, and corneal infections, can occur, potentially leading to vision loss.

### 9.2.6 Vertical Transmission

While data regarding the effects of MPox during pregnancy is scarce, case studies have indicated the possibility of vertical transmission via the placenta, which can result in fetal death and stillbirths. A meta-analysis has further underscored a significant association between MPox infection during pregnancy and perinatal loss, primarily attributed to the heightened risk of vertical transmission.

### 9.2.7 Prevention and Treatment

As of March 22, 2023, WHO and CDC guidelines do not prescribe targeted therapy for MPoxV. Nevertheless, specific antiviral drugs like Cidofovir, Tecovirimat, and Brincidofovir, along with the JYNNEOS vaccine, have shown efficacy as prophylactic measures. It's important to note that these vaccines are supplied at no cost but are intended for use in high-risk individuals rather than mass vaccination. SPox vaccines like MVA-BN and ACAM-2000 are used to guard against MPox. These vaccines have demonstrated approximately 85% protection against MPoxV. Considering recent evidence suggesting sexual transmission during the outbreak, it is advisable to implement contact tracing, population surveillance, and targeted ring vaccination. Genomic sequencing and PCR assays are available for virus detection. To effectively control the outbreak, WHO endorses strategies that include raising general awareness, conducting relevant diagnostic testing, identifying, and isolating affected individuals, regulating wild animal trade, and promoting the safe utilization of thoroughly cooked meat.

## 9.3 Web-Based Information on MPox

Natural language processing has become a powerful tool for curating web-based information on specific topics. For instance, Kolluri et al. [68] introduced the 'POXVERIFI' browser extension, using the (bidirectional encoder representations from transformers) BERT technique. By allowing users to install the extension, rate articles, and provide crowd-sourced votes were used to generate precise labels for new sources, contributing to the measurement of MPox-related misinformation.

In a similar vein, [106] developed a hybrid CNN–LSTM model for classifying MPox-related Twitter posts. Their model demonstrated an exceptional precision rate of 94.00% in evaluating the reactions of Twitter users toward MPox. Independent ML techniques were also applied to analyze 352,182 Twitter posts related to MPox in another study [107]. These techniques effectively clustered topics into themes that included the marginalization of minority groups, safety apprehensions, and the erosion of trust in institutions. Transformers like BERT and BERTopic played a pivotal role in facilitating sentiment analysis. Furthermore, in a study focusing on 15,936 MPox-related tweets in German, a mixed-methods research approach was employed in conjunction with ML techniques [108]. The authors underscored the necessity of adopting a multidisciplinary strategy to mitigate and prevent MPox-related misinformation.

## 10 Ethical Concerns of AI for Patient Care

Integration of DL into patient care demands meticulous attention to ethical considerations:

- **Privacy and Data Security**: It's essential to ensure the safeguarding of extensive health records available to AI systems. Preventing unauthorized entry, breaches, or abuse is vital. Employing techniques like data anonymization, encryption, and robust access controls becomes necessary to protect patient confidentiality.
- **Regulatory and Ethical Framework**: Achieving compliance with applicable laws, regulations, and ethical standards is a cornerstone in the integration of AI in patient care.
- **Unforeseen Outcomes**: The integration of DL into healthcare may yield unforeseen consequences, including reduction of skills or reduced human oversight, which might impact the quality of care. To mitigate risks and ensure positive outcomes, adequate education of healthcare staff in AI tool usage and continuous monitoring and evaluation of DL systems are essential [120].

## 11 Future Directions

Key directions encompass modernizing management, leveraging GAN for augmentation, embracing DL advancements, exploring novel architectures, strategic pandemic planning, and advancing CADx development for MPox healthcare advances, as illustrated in Figure 16.

### 11.1 Modernizing Healthcare Management

In the realm of MPox management, there's the possibility for a secure, cloud-based system. This system could integrate patient information, portable device information, mobile data, and cloud-based AI models. Healthcare professionals would remotely access these resources, including uploading skin lesion images for diagnosis. Furthermore, it could encompass clinical, laboratory, epidemiological, and demographic parameters, allowing system-wide healthcare resource allocation analysis. To curb infection spread, integrated teleconsultation via remote video conferencing could enable remote dispensing of symptomatic treatment for mild cases. This scalable system could aggregate data from multiple locations, aiding in trend detection for strategic planning. Sensitivity analysis can quantify the impact of diverse datasets on the effectiveness of classifiers.

### 11.2 Augmentation using GAN

MPox, a recently identified infectious disease, faces a challenge due to the scarcity of publicly available labeled datasets. To address this limitation, the dataset size is augmented through on-the-fly data augmentation. However, future research aims to enhance training sets by generating synthetic examples using GANs.

### 11.3 Novel DL Architectural Ideas

The novel infectious disease will share certain common characteristics with MPox. Consequently, there is potential to apply deep CNN and ViT concepts to diagnose this new variant of MPox.

- The Channel boosting based on the integration of CNNs and ViT at abstract and target levels captures minor contrast and texture variations in the infected region. Boosting extracts global structural variations through diverse channel-wise information exploitation.
- Boosting can be achieved through TL that leverages pre-trained network knowledge, fine-tuning for new infectious challenges [121].
- The deep features ensemble of deep CNNs and ViT can effectively learn the infection patterns specific to the new infectious disease, distinguishing them from established MPox patterns and deformed regions.

### 11.4 Development of CADx

The researcher could concentrate on developing a CADx system and a fully automated pipeline to establish a foundation for upcoming investigations. This approach enables cross-validation of newly developed models with the current dataset, streamlining the workload and lessening reliance on medical experts. Additionally, it aids doctors in decision-making regarding new datasets. The prompt and computer-aided diagnosis facilitated by the DL contributes to early detection, saving lives, and thereby generating a positive societal impact.

### 11.5 DL Advancements in Healthcare

DL is driving a profound transformation in healthcare, revolutionizing disease detection, management, and treatment, bolstering the healthcare system. AI often outperforms conventional methods. In the future, AI's potential applications extend to the drug industry, facilitating the finding of more powerful therapeutic compounds for various ailments. AI has also enhanced patient care and empowered clinicians to deliver precise and timely diagnoses, significantly reducing detection time and bridging the gap between detection and treatment. It equips healthcare providers with the capability to adapt and refine ongoing treatment plans derived from previously archived patient information. Shortly, DL will enhance the ability to diagnose, identify risk factors, and plan treatment, offering substantial enhancements in patient outcomes and operational efficiency.

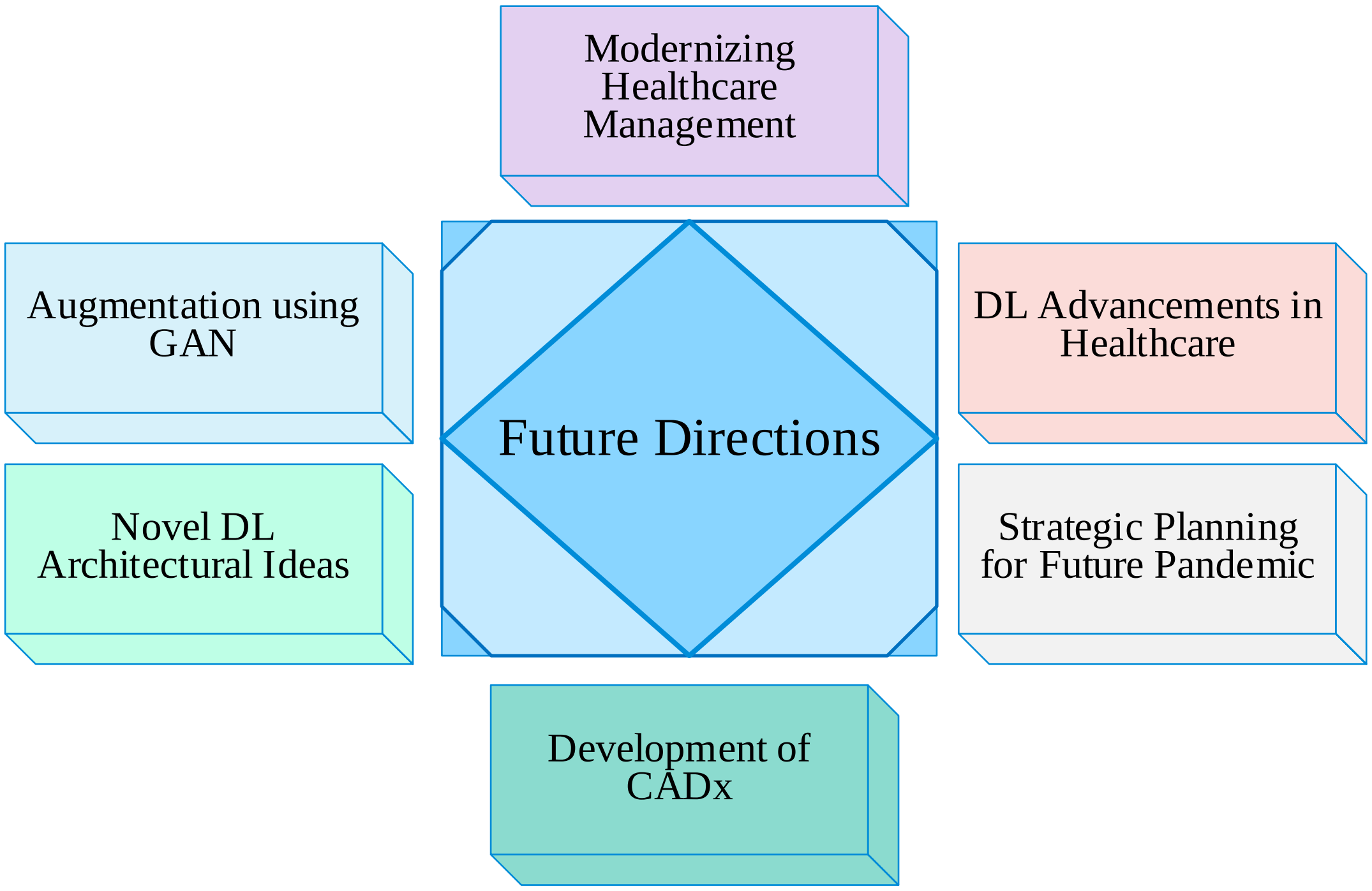

Figure 16: Future Directions Mpox Detection.

Likewise, as the world grapples with the ongoing appearance of infectious diseases, the adaptability of DL techniques becomes paramount. DL algorithms must evolve to meet the challenges posed by highly variable and globally distributed infectious diseases effectively. Nevertheless, it's important to recognize that DL has its constraints, for example, data and technical inefficiencies, which currently hinder its full-scale implementation in healthcare. Addressing these challenges will be pivotal in realizing the full potential of DL in healthcare.

### 11.6 Strategic Planning for Future Pandemic

Lessons drawn from previous outbreaks, notably the experience with COVID-19, can guide more efficient responses to both short-term and long-term viral outbreaks. These past outbreaks have led to the establishment of improved surveillance networks, accelerated development of therapeutic solutions, and increased funding to address such crises. The refinement of control strategies, ongoing research, and comprehensive studies across various facets of disease systems can augment public health responses in the years ahead. Despite the continued threat of outbreaks, the application of technological tools and concerted efforts can result in the development of effective intervention strategies before the occurrence of the next major event.

## 12 Conclusion

The current MPox outbreak has raised global concerns, emphasizing the need for preparedness, analysis, and treatment. Recent years have witnessed significant advancements in DL within health science research and its practical applications. Our analysis covers the use of DL models for MPox detection, and outbreak prediction, and delves into DL's role in infectious disease diagnosis. Currently, the rapid and integral element in the diagnosis and treatment pipeline of infectious MPox diseases is the use of deep CNNs. These networks leverage and identify characteristic patterns indicative of irregularities and abnormalities in medical images of infectious diseases. This comprehensive review offers an extensive survey of DL methodologies, with a specific emphasis on analyzing MPox viral infections within skin lesion images. The survey categorizes and assesses the efficacy of utilization of deep CNNs, deep CNN ensemble and hybrid learning, integration of newly developed techniques, and ViT. Additionally, the review provides an in-depth overview of each study, including details on dataset characteristics and pre-processing, class numbers, data distribution, and performance evaluation criteria. Furthermore, the paper sheds light on diagnostic instruments, accessible data resources, limitations, and the challenges encountered during the epidemic. This study delivers valuable insights for DL and medical image research, intending to advance the establishment of a standardized and comprehensive system for MPox detection. Moreover, DL models demonstrate proficiency in handling extensive medical datasets, facilitating the prediction of future disease occurrences and their probabilities. Additionally, a key application of DL in healthcare is the prediction of personalized treatment plans based on individual patient characteristics and treatment preferences.

**Acknowledgment**

We thank the Artificial Intelligence Lab, Department of Computer Systems Engineering, University of Engineering and Applied Sciences (UEAS), Swat, for providing the necessary resources.

**Conflicts of interest**: The authors declare that they have no known competing financial interests or personal relationships that could have appeared to influence the work reported in this paper.

**Institutional Review Board Statement**

Not applicable.

**Informed Consent Statement**

Not applicable.

**Data Availability Statement** Correspondence and requests for materials should be addressed to Saddam Hussain Khan.

**Full Forms and Abbreviations Key**

| Full Form | Abb. | Full Form | Abb. |
|---|---|---|---|
| Accuracy | Acc | Support Vector Machine, | SVM. |
| Area under Curve | AUC | Naïve Bayes | nB. |
| Balanced Accuracy | BA | Random Forest | RF. |
| Convolutional Neural Network | CNN | Hand-foot-mouth disease | HFMD. |
| Cross-Validation | Cv | Bagging | B |
| Data Augmentation | DA | bidirectional long/short-term memory | BiLSTM |
| Data Organization | Do | Ensemble | En |
| Deep Learning | DL | Monekpox | MPox |
| Dermatologist | D | Local Interpretable Model Agnostic Explanations | LIME. |
| Feature Selection | FS | Decision Tree | DT. |
| Feature Selection | FS | Rotation Forest | RTF |
| Grey Wolf Optimizer Algorithm | GWO | Al-Biruni Earth radius optimization-based stochastic fractal search | BERSFS. |
| Machine Learning | ML | Matthew's Correlation Coefficient. | MCC |
| Normalization | Nor | Long-short term memory | LSTM |
| Partitioning algorithm | PA | Vision Transformer | ViT. |
| Precision | P | K-nearest neighbor, | KNN. |
| Recall | R | Harris Hawks Optimizer | HHO. |
| Resizing | Rs | Confidence Interval | CI. |
| Sensitivity | Sen | Metaheuristics optimization-based weighted average En model | MO-WAE |
| Specificity | Sp | Metaheuristic Optimization | MO |
| Splitting | Splt | Random Subspace | RS |
| Transfer learning | TL | Image | Img. |